# Detailed illustration of accuracy of presently used nuclear-mass models

A. Sobiczewski[*,a,b,c], Yu.A. Litvinov[c,1], M. Palczewski[a,2]

[a]*National Centre for Nuclear Research, Hoża 69, 00-681 Warsaw, Poland*
[b]*Joint Institute for Nuclear Research, 141980 Dubna, Russian Federation*
[c]*GSI Helmholtzzentrum für Schwerionenforschung GmbH, 64291 Darmstadt, Germany*

**Abstract**

The accuracy of description of measured nuclear masses by presently used nuclear-mass models is studied. Twelve models of various kinds are considered, eleven of the global character and one local model specially adapted to description of heavy nuclei. To decrease the number of nuclei over which the accuracy is averaged, the global region ($Z, N \geq 8$) is divided into four subregions, in which the accuracy is studied separately. Still, to reach the best precision, the accuracy is finally investigated without any averaging, for each nucleus separately. The results are presented in a form of colored maps, large enough to be easily and accurately read.

Besides the accuracy of the studied models, also their predictive power is considered.

It is found that the accuracy of description of mass strongly depends on a nuclear-mass model and on the region of nuclei to which the model is applied. The best accuracy is obtained by the recent two Chinese models WS3+ and WS4+. Generally, no clear, strong correlation between the accuracy of description of already known masses by a given model and its predictive power for new masses is observed. Still, such correlation is found for separate models and in separate regions of nuclei. More often for the macroscopic-microscopic models than for the other approaches.

[*]Corresponding author:
  *Email address:* `E-mail:adam.sobiczewski@fuw.edu.pl`; tel.:+48 22 6216085 (A. Sobiczewski)
[1]E-mail: y.litvinov@gsi.de
[2]E-mail: michal.palczewski@ncbj.gov.pl

**Contents**

## 1. Introduction

Nuclear masses are basic quantities for understanding many properties of atomic nuclei and also of nuclear processes in nuclear physics and astrophysics. Due to this, there are continuing big efforts in measuring them for more and more nuclei (some of them being very exotic ones), in increasing the accuracy of the measurements, but also in understanding the role of this progress in nuclear physics and astrophysics [1–18].

On the theoretical side, there are big efforts in constructing more and more subtle models trying to reproduce better and better measured masses and clear up the relation of them with nuclear interactions and the structure of a nucleus. Also the efforts in the elaboration of theoretical methods using data on masses for the extension of our knowledge of nuclear structure and nuclear processes [19–25].

From time to time, the progress in the measurements is summarized by an evaluation of the measured values. The last evaluation was completed recently [26], about a decade after the previous one [27]. In comparison to the number of nuclei with experimentally known masses in 2003 (2226), this number (2436) has now increased by 210 nuclei. This large number of new masses gave us an exceptional opportunity to perform an important test of the accuracy of various mass models. The test was done in our recent papers [28–30].

The objective of the present study is to extend this analysis in two respects. One is the inclusion to the consideration of two recent models [31, 32], not considered in our previous studies. The second respect is probably even more important. It was shown in Ref. [29] that the accuracy of a given model strongly depends on the nuclei to which it is applied. Thus, any averaging of the accuracy over a number of nuclei decreases the quality (precision) of the information on the accuracy of the model. In particular, the averaging over the whole (global) considered region (usually over the nuclei with $Z, N \geq 8$, where $Z$ and $N$ are proton and neutron numbers, respectively), which is usually done, leads to especially poor information, although still useful. Due to this, besides the accuracy averaged over various regions of nuclei, we present in this paper the accuracy of each of the considered models without any averaging, i.e. for each of the nuclei separately. This is done in the form of maps large enough to be read with a reasonable accuracy. Thus, in this form, the information is rather concise and gives the opportunity for a fast orientation for which nuclei the model is more accurate and for which less. An example of such illustration was presented in Ref. [33]. It was done, however, only for the region of heavy nuclei ($Z \geq 82$) and only for seven models. This paper extends the illustration to the global region divided, however, to four subregions. Twelve models of various kinds are considered, from relatively old ones up to very recent. Besides supplying a general knowledge on the accuracy of the present theoretical description of masses, the results of the analysis may provide useful information for users of the models to help in choosing the most appropriate one, when studying specific nuclei. For example, the choice of the proper model was crucial for the successful prediction of the properties of the decay chains of the nuclei of the as-yet-unobserved element $Z = 117$ [34]. A comparison of the predicted and experimentally obtained values was given in Ref. [35] describing the discovery of this element. Additionally, the results of the present study may also be helpful for the authors of the models in improving them.

The predictive power of the considered models is also studied in the paper.

## 2. Mass models considered

As already stated in the Introduction, twelve models are discussed; from relatively old, but still used ones: FRDM [36] and DZ [37], up to very recent: INM [31] and WS4+ [32]. The models are of different nature. Seven of them are of the macroscopic-microscopic type, two of the purely microscopic (self-consistent) nature and three of still other kind. The macroscopic-microscopic models are: the Warsaw local Heavy-Nuclei (HN) model, specially adapted to describe heavy nuclei [38] (see also Ref. [39]), FRDM model, the FRDM12 model [40], the nuclear Thomas-Fermi model (TF) [41], the Lublin-Strasbourg drop (LSD) [42], and the recent Weizsäcker-Skyrme models: WS3+ [21] and WS4+ [32]. Important for the latter two models is the use of the radial basis function (RBF) method (see Ref. [21], where the method is described in detail). Here, the notation WS3+ and WS4+ is a shortened one for the WS3+RBF and WS4+RBF, respectively. The two purely microscopic models are the Hartree-Fock-Bogoliubov mean field models: one with the BSk21 Skyrme interaction (HFB21) [19], and the other with the D1M Gogny forces (GHFB) [43]. The last three models of other kinds are those of Duflo and Zuker (DZ) [37] (see also Ref. [20]), of Koura et al. (KTUY) [44] and of Nayak and Satpathy (INM) [31]. The DZ and the KTUY models use a large number of parameters directly adjusted to experimental masses. The DZ model uses 28 and the KTUY 34 parameters.

The models discussed presently are not the same as studied previously [29, 30]. With respect to that set, two models: FRLDM [36] and WS3.6 [22] are skipped and four other: FRDM12 [40], HN [38], INM [31] and WS4+ [32] are added.

## 3. Accuracy of the models averaged over various regions of nuclei

The results presented in this section are an extension of those of Ref. [29]. The set of the models considered presently differs by five models from those used in Ref. [29], as stated earlier. Also, one of the new models (WS4+) differs from all others in the respect that it was adjusted to the recent data [26], while all others were fitted to earlier evaluations.

This should be kept in mind when discussing the results. Thus, a better accuracy of its results may come, to a large extent, from this fact and not only from a better, more physical nature of that model.

Table A gives the accuracy of the description of the experimental masses by each of the models for six regions of nuclei: global, light, medium-I, medium-II, heavy and the heaviest nuclei. The accuracy is expressed by the rms (root-mean-square) values of the discrepancies between the calculated and experimental masses. The average values of the discrepancies, $\bar{\delta}$, are also given. Both rms and $\bar{\delta}$ are in MeV. As mentioned above, experimental masses are taken from the recent evaluation [26]. For each region and each model, the number of nuclei with both calculated and evaluated masses in 2012, $N_{\text{nucl}}$, are also shown. They provide the information on how many masses are involved in the description. For each model, the year of its publication is shown as well.

Table A
The rms and the average value, $\bar{\delta}$, of the discrepancies calculated for the global ($Z, N \geq 8$), light ($8 \leq Z < 28, N \geq 8$), medium-I ($28 \leq Z < 50$), medium-II ($50 \leq Z < 82$), heavy ($Z \geq 82$) and the heaviest ($Z \geq 100$) nuclei, obtained with the use of the specified models, are given. The numbers of nuclei with both calculated and evaluated in 2012 masses, $N_{\text{nucl}}$, are also shown. For the HN model, the heavy-nuclei region is specified as: $Z \geq 82$, $N \geq 126$

| Model | LSD | FRDM | FRDM12 | TF | HFB21 | GHFB | DZ | KTUY | INM | WS3+ | WS4+ | HN |
|---|---|---|---|---|---|---|---|---|---|---|---|---|
| Year | 2003 | 1995 | 2016 | 1996 | 2010 | 2009 | 1995 | 2005 | 2012 | 2010 | 2014 | 2001 |
| No. | 1 | 2 | 3 | 4 | 5 | 6 | 7 | 8 | 9 | 10 | 11 | 12 |
| Global | | | | | | | | | | | | |
| $N_{\text{nucl}}$ | 2316 | 2353 | 2353 | 2351 | 2353 | 2353 | 2353 | 2353 | 2353 | 2353 | 2353 | |
| rms | 0.608 | 0.654 | 0.579 | 0.649 | 0.572 | 0.789 | 0.394 | 0.701 | 0.362 | 0.248 | 0.170 | |
| $\bar{\delta}$ | -0.027 | -0.059 | -0.010 | 0.027 | 0.030 | -0.103 | -0.032 | -0.058 | -0.011 | -0.008 | 0.000 | |
| Light | | | | | | | | | | | | |
| $N_{\text{nucl}}$ | 332 | 335 | 335 | 335 | 335 | 335 | 335 | 335 | 335 | 335 | 335 | |
| rms | 1.046 | 1.144 | 1.056 | 1.054 | 0.911 | 1.087 | 0.546 | 0.692 | 0.502 | 0.362 | 0.247 | |
| $\bar{\delta}$ | -0.180 | -0.162 | -0.059 | -0.012 | 0.035 | -0.365 | 0.058 | -0.055 | 0.034 | 0.063 | 0.004 | |
| Medium-I | | | | | | | | | | | | |
| $N_{\text{nucl}}$ | 574 | 575 | 575 | 575 | 575 | 575 | 575 | 575 | 575 | 575 | 575 | |
| rms | 0.650 | 0.664 | 0.618 | 0.701 | 0.578 | 0.748 | 0.406 | 0.783 | 0.422 | 0.277 | 0.175 | |
| $\bar{\delta}$ | 0.042 | 0.072 | 0.010 | 0.222 | 0.099 | 0.30 | -0.097 | -0.313 | -0.102 | -0.055 | -0.005 | |
| Medium-II | | | | | | | | | | | | |
| $N_{\text{nucl}}$ | 961 | 970 | 970 | 970 | 970 | 970 | 970 | 970 | 970 | 970 | 970 | |
| rms | 0.451 | 0.475 | 0.368 | 0.501 | 0.455 | 0.556 | 0.328 | 0.542 | 0.306 | 0.207 | 0.148 | |
| $\bar{\delta}$ | -0.101 | 0.132 | -0.042 | -0.209 | -0.034 | -0.032 | -0.026 | 0.216 | -0.008 | -0.006 | 0.005 | |
| Heavy | | | | | | | | | | | | |
| $N_{\text{nucl}}$ | 449 | 473 | 471 | 473 | 473 | 473 | 473 | 473 | 473 | 473 | 473 | 312 |
| rms | 0.349 | 0.448 | | 0.444 | 0.458 | 0.971 | 0.376 | 0.869 | 0.254 | 0.179 | 0.133 | 0.355 |
| $\bar{\delta}$ | 0.156 | 0.006 | 0.066 | 0.302 | 0.073 | -0.227 | -0.032 | -0.312 | 0.061 | -0.004 | -0.009 | -0.118 |
| Heaviest | | | | | | | | | | | | |
| $N_{\text{nucl}}$ | 32 | 36 | 36 | 34 | 36 | 36 | 36 | 36 | 36 | 36 | 36 | 36 |
| rms | 0.227 | 0.676 | 0.690 | 0.439 | 0.322 | 1.127 | 0.828 | 1.236 | 0.471 | 0.126 | 0.130 | 0.118 |
| $\bar{\delta}$ | -0.135 | -0.490 | -0.610 | 0.411 | -0.014 | 0.957 | -0.233 | 1.200 | 0.284 | 0.052 | -0.021 | -0.026 |

One can see in the table that for each region of the nuclear chart, the rms changes quite strongly from one model to another. For example, for the light-nuclei region, the rms changes from 0.247 MeV (WS4+) to 1.144 MeV (FRDM). For a given model, the rms also strongly changes with the change of the region. For example, the rms changes from 0.448 MeV (heavy) to 1.144 MeV (light), for the FRDM approach. Also the average discrepancy $\bar{\delta}$ strongly depends on the model (e.g. it changes from -0.004 MeV for WS3+ to 0.302 MeV for TF in the heavy region) and also on the region for a given model (e.g. it changes from -0.012 MeV in the light to 0.302 MeV in the heavy region for the TF model).

The dependence of the rms discrepancy on the model and on the region of nuclei is illustrated in Figs. 1 and 2. One can see in Fig. 1 that the dependence of the rms on the region of nuclei is very strong, especially for the FRDM, TF and LSD models. For these three approaches, not only the dependence, but also the values of the rms are very close to each other for all the regions investigated. This reflects the fact that the nature of these macroscopic-microscopic models is very similar. The dependence of the rms of the WS3+ and WS4+ models is also similar to each other, also reflecting the similarity of their nature. Similar also is the dependence of the INM model although its nature is rather different. The values of the latter three models, however, are much smaller than those for the FRDM, TF and LSD models. It is worthwhile noting, that the rms of all these five models systematically decrease with increasing mass of the nuclei. In other words, the quality of the description of masses systematically increases when one passes from lighter to heavier nuclei. This might be interpreted to mean that the assumption of a good mean field, on which all the models are based, is better fulfilled for heavier nuclei.

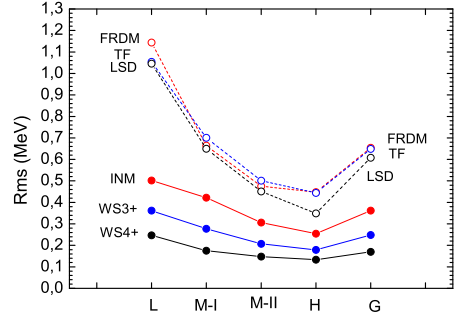

**Fig. 1**:

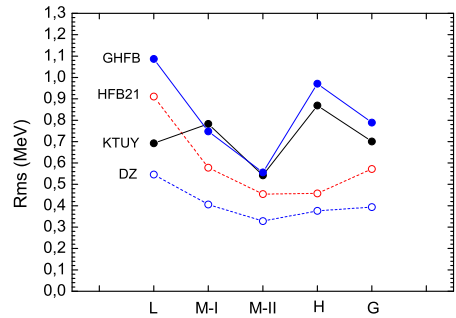

**Fig. 2**:

Similar behaviour is observed in Fig. 2 for the HFB21 and DZ approaches. However, the results for the GHFB and KTUY models show different trends. Here, it is interesting to note that the HFB21 and GHFB models, although using the same approach (HFB), they show a difference in the dependence of the rms on the region of nuclei. This is probably the effect of the difference in the effective forces used, but also of a difference in the ways of introducing the correlations.

This illustrates the sensitivity of the dependence of the rms on the region of nuclei to the details of a particular model.

To see the dependence of the average accuracy of the description of measured masses on a model and on the region of nuclei, let us show the rms as a function of a model in three regions of nuclei: M-II (i.e. medium-II), H (heavy) and the region of heaviest nuclei. The results are shown in Figs. 3, 4 and 5, respectively. A given model is identified in these figures by the number prescribed to it in Table A.

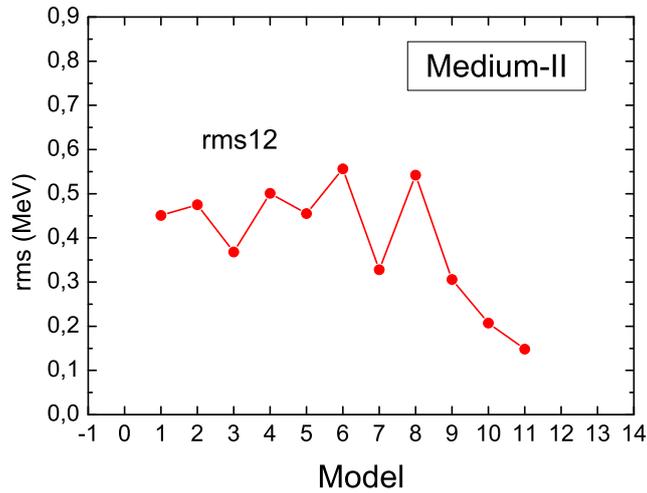

Fig. 3:

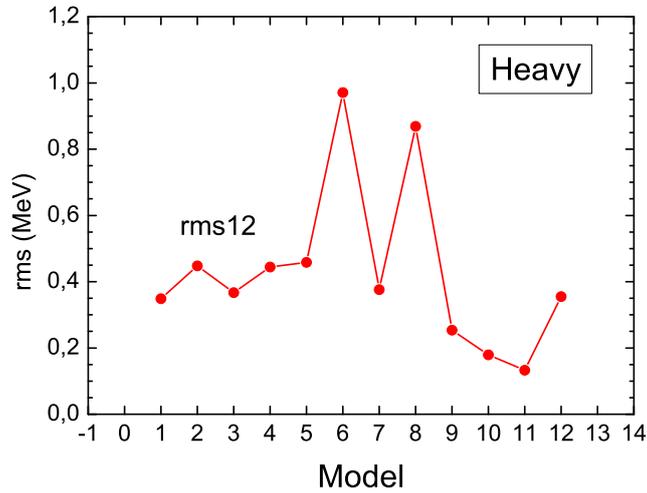

Fig. 4:

It is seen in Fig. 3 that the best description of masses in the M-II region is obtained in the case of the macro-micro models WS3+ and WS4+. A relatively good description is also observed for the INM, DZ and FRDM12 approaches. In the heavy-nuclei region (Fig. 4), the best description is observed again for the two macro-micro approaches WS3+ and WS4+. A relatively good accuracy is also seen in the case of the INM, LSD, FRDM12 and DZ models. Finally, in the region of the heaviest nuclei (Fig. 5), the best result is observed in the case of the three macro-micro models: WS3+, WS4+ and HN. The latter result is important for people predicting the properties of super-heavy nuclei. A relatively good description is also seen for the (again macro-micro) approach LSD.

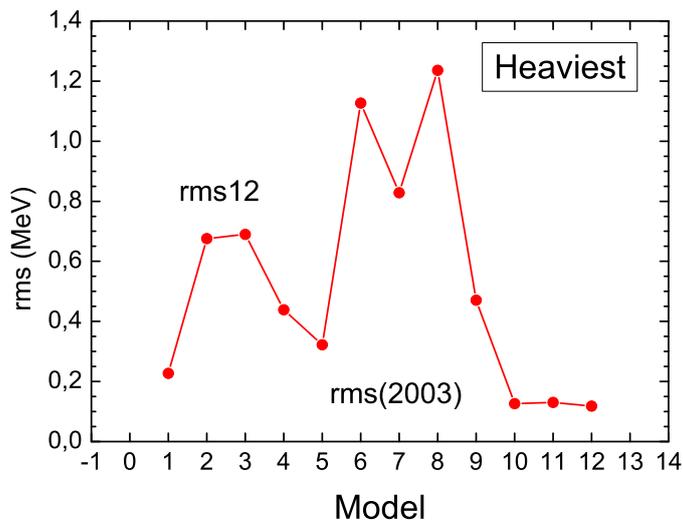

**Fig. 5**:

In conclusion of these three illustrations, one can ask an important question: what can be the reason for the best description of masses by the macro-micro models? A natural hypothesis would be that just the macroscopic part (usually the liquid drop) of these approaches is probably a physically good model for description of a number of nuclear properties, among them nuclear mass, averaged over microscopic effects.

### 4. Predictive power of the models

The results presented in this section are an extension of those of Ref. [30]. The main difference is that the set of the models considered presently differs by four models (INM and WS4+ models instead of FRLDM and WS3.6). Also, one of the new models (WS4+) differs from all others in the respect that it was adjusted to the recent data [26], while all others were fitted to earlier evaluations. Thus, the predictive power discussed for this model is not a real ability to predict masses not known yet. Still, it is interesting to see how this different situation of this model will be reflected in the results.

The main subject of Ref. [30] was to study the relation between two important properties of a given model: its accuracy of description of already measured masses and its predictive power for masses not-yet-known. One might expect that a model which describes well masses already known contains a good physics inside it and, thus, will be able to predict properly masses not-yet-presently known. Really, for predicting not-yet-known masses, people usually use models which best describe masses already known. In other words, they assume a strong correlation between a good accuracy of description of known masses by a model with a good prediction by it of new masses. The result of the paper [30] is that, generally, a clear, strong such correlation does not appear in the presently used models.

Let us test this result for ten nuclear models considered in Table B, where the respective results are presented.

Table B
Two sets of the rms values are presented: rms(2003) and rms(new). The values are given separately for the global $(Z, N \geq 8)$, light $(8 \leq Z < 28, N \geq 8)$, medium-I $(28 \leq Z < 50)$, medium-II $(50 \leq Z < 82)$, and heavy $(Z \geq 82)$ regions of nuclei. Also shown are the corresponding numbers of used nuclei: $N_{\text{nucl}}(2003)$ and $N_{\text{nucl}}(\text{new})$.

| Model | LSD | FRDM | TF | HFB21 | GHFB | DZ | KTUY | INM | WS3+ | WS4+ |
|---|---|---|---|---|---|---|---|---|---|---|
| (Year) | (2003) | (1995) | (1996) | (2010) | (2009) | (1995) | (2005) | (2012) | (2010) | 2014 |
| No. | 1 | 2 | 3 | 4 | 5 | 6 | 7 | 8 | 9 | 10 |
| Global | | | | | | | | | | |
| $N_{\text{nucl}}(2003)$ | 2127 | 2134 | 2134 | 2134 | 2134 | 2134 | 2134 | 2134 | 2134 | 2134 |
| $N_{\text{nucl}}(\text{new})$ | 189 | 219 | 217 | 219 | 219 | 219 | 219 | 219 | 219 | 219 |
| rms(2003) | 0.620 | 0.655 | 0.638 | 0.578 | 0.799 | 0.358 | 0.651 | 0.299 | 0.217 | 0.202 |
| rms(new) | 0.627 | 0.765 | 0.805 | 0.646 | 0.764 | 0.673 | 1.092 | 0.712 | 0.374 | 0.155 |
| Light | | | | | | | | | | |
| $N_{\text{nucl}}(2003)$ | 319 | 319 | 319 | 319 | 319 | 319 | 319 | 319 | 319 | 319 |
| $N_{\text{nucl}}(\text{new})$ | 13 | 16 | 16 | 16 | 16 | 16 | 16 | 16 | 16 | 16 |
| rms(2003) | 1.068 | 1.154 | 0.990 | 0.933 | 1.053 | 0.543 | 0.731 | 0.487 | 0.326 | 0.327 |
| rms(new) | 1.236 | 1.558 | 1.923 | 1.021 | 1.782 | 0.889 | 1.092 | 0.680 | 0.579 | 0.242 |
| Medium-I | | | | | | | | | | |
| $N_{\text{nucl}}(2003)$ | 508 | 508 | 508 | 508 | 508 | 508 | 508 | 508 | 508 | 508 |
| $N_{\text{nucl}}(\text{new})$ | 66 | 67 | 67 | 67 | 67 | 67 | 67 | 67 | 67 | 67 |
| rms(2003) | 0.669 | 0.679 | 0.725 | 0.613 | 0.800 | 0.363 | 0.643 | 0.273 | 0.213 | 0.220 |
| rms(new) | 0.654 | 0.721 | 0.715 | 0.529 | 0.466 | 0.649 | 1.368 | 0.950 | 0.457 | 0.161 |
| Medium-II | | | | | | | | | | |
| $N_{\text{nucl}}(2003)$ | 894 | 895 | 895 | 895 | 895 | 895 | 895 | 895 | 895 | 895 |
| $N_{\text{nucl}}(\text{new})$ | 67 | 75 | 75 | 75 | 75 | 75 | 7 | 75 | 75 | 75 |
| rms(2003) | 0.445 | 0.461 | 0.481 | 0.439 | 0.549 | 0.300 | 0.543 | 0.256 | 0.193 | 0.155 |
| rms(new) | 0.533 | 0.598 | 0.676 | 0.611 | 0.617 | 0.567 | 0.532 | 0.647 | 0.309 | 0.151 |
| Heavy | | | | | | | | | | |
| $N_{\text{nucl}}(2003)$ | 406 | 412 | 412 | 412 | 412 | 412 | 412 | 412 | 412 | 412 |
| $N_{\text{nucl}}(\text{new})$ | 43 | 61 | 59 | 61 | 61 | 61 | 61 | 61 | 61 | 61 |
| rms(2003) | 0.343 | 0.401 | 0.442 | 0.416 | 1.000 | 0.279 | 0.797 | 0.211 | 0.159 | 0.134 |
| rms(new) | 0.407 | 0.673 | 0.467 | 0.676 | 0.744 | 0.749 | 1.250 | 0.436 | 0.263 | 0.121 |

The rms(2003) values are the average values of the discrepancies between the calculated masses and the experimental ones evaluated in 2003 [27], while the rms(new) correspond to the new masses evaluated in 2012 [26], which were not known in 2003. All the data are given for five regions of nuclei specified in the table caption. For each region and each model, the number of nuclei with both calculated and evaluated masses in 2003, $N_{\text{nucl}}$ (2003), and new ones, $N_{\text{nucl}}(\text{new})$, are also shown. They give us the information how many new masses are involved in the description, in comparison to masses known in 2003. For each model, the year of its publication is shown as well.

It is also seen in the table that both, rms(2003) and rms(new), depend strongly on the model for a given region of nuclei, and on the region of nuclei for a given model. Let us illustrate both of them. They are better seen when presented in a graphical form. The illustration is done in another, more direct way than in the study [30].

*4.1. Dependence on the model for a given region of nuclei*

The dependence of the rms(2003) and the rms(new) on the model for a given region of nuclei, is illustrated in figures 6 to 10 for the five regions of nuclei: global, light, medium-I, medium-II and heavy regions. A nuclear model is identified by the number prescribed to it in Table B.

One can see in Fig. 6 that in the global region, a good predictive power, defined by the condition:

$$rms(new) \approx rms(2003) \qquad (1)$$

is observed only for the LSD, HFB21, GHFB and WS4+ models, while the lowest value of both rms is obtained for the macro-micro WS4+ model. The condition (1) is especially badly fulfilled by the KTUY approach.

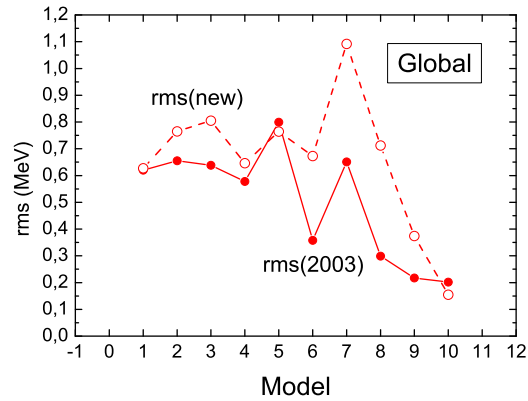

**Fig. 6**:

For the region of light nuclei (Fig. 7), a good predictive power is obtained only for the HFB21 and WS4+ models. This property is especially badly fulfilled by the TF and GHFB approaches.

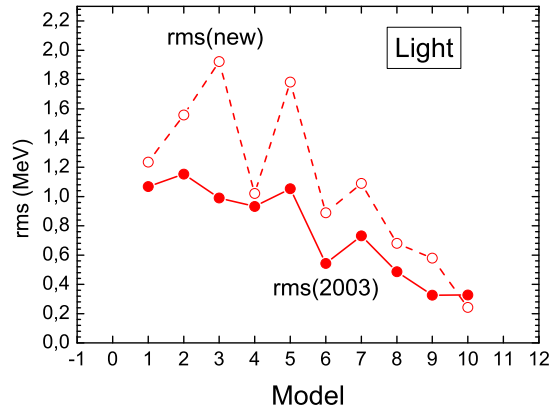

**Fig. 7**:

In the medium-I region (Fig. 8), a good predictive power is observed for the LSD, FRDM, TF and WS4+ models. This property is especially badly fulfilled by the KTUY model.

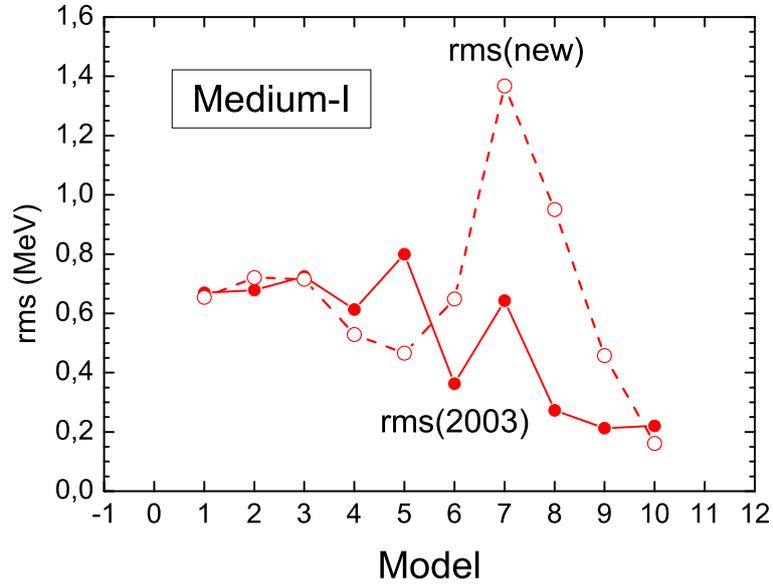

**Fig. 8**:

For the region of medium-II nuclei (Fig. 9), a good predictive power is only observed for the KTUY and WS4+ approaches. This property is especially badly fulfilled by the INM model.

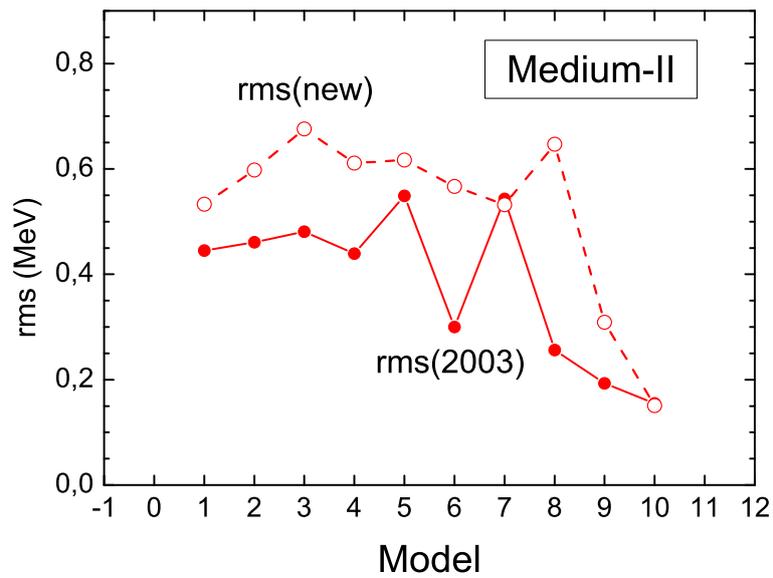

**Fig. 9**:

Finally, in the heavy-nuclei region (Fig. 10), a good predictive power is obtained for the LSD, TF and WS4+ models, and it is especially badly fulfilled by the FRDM, KTUY and INM approaches.

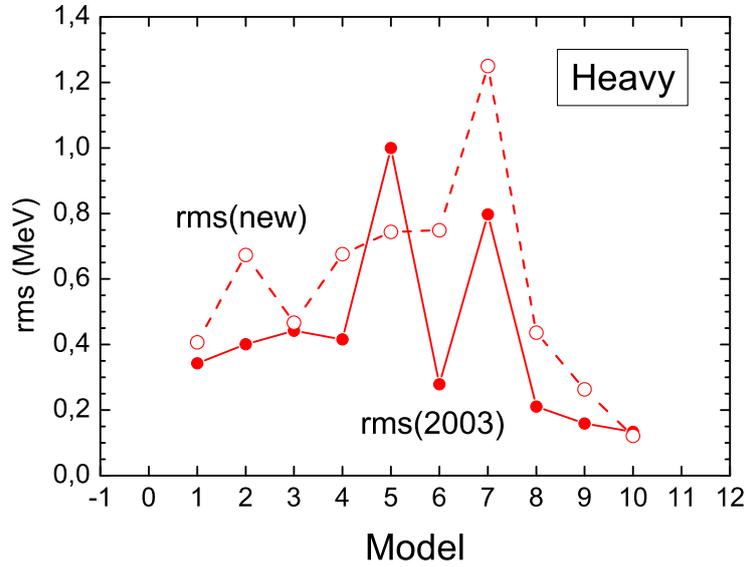

Fig. 10:

## 4.2. Dependence on the region of nuclei for a given model

The dependence of rms(2003) and rms(new) on the region of nuclei for a given model is illustrated in figures 11 to 20, i.e. for ten models discussed in this chapter. In the figures, L, M-I, M-II, H, and G denote the light, medium-I, medium-II, heavy, and global regions of nuclei, respectively.

For the LSD model (Fig. 11), a relatively good predictive power is obtained for all regions of nuclei. Especially good, it is observed for the M-I and G regions.

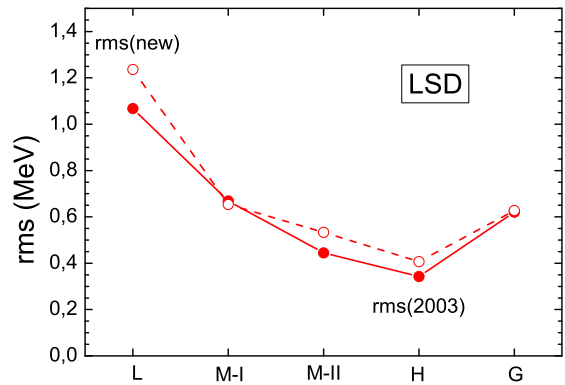

Fig. 11:

For the FRDM model (Fig. 12), a good predictive power is obtained only for the M-I region and especially bad for the L and H regions.

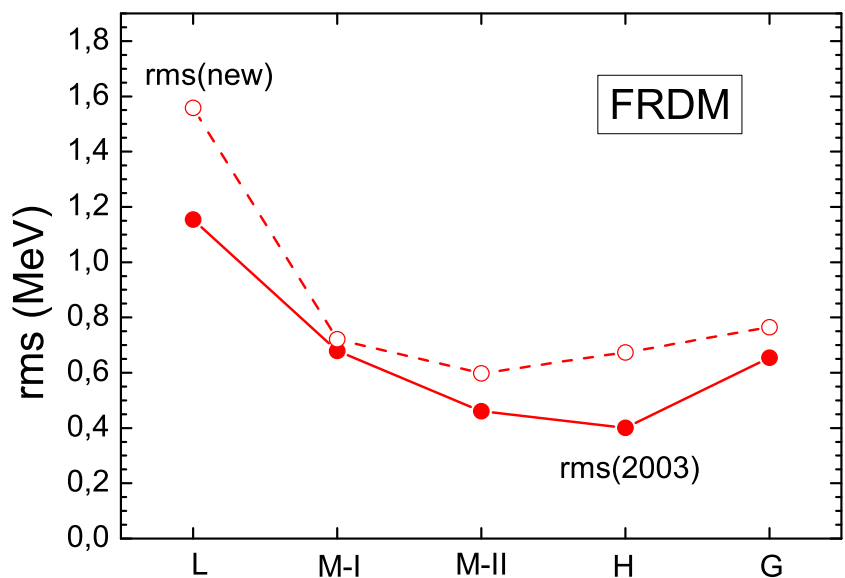

**Fig. 12**:

For the TF model (Fig. 13), a good predictive power is observed for the M-I and H regions and especially bad for the L region.

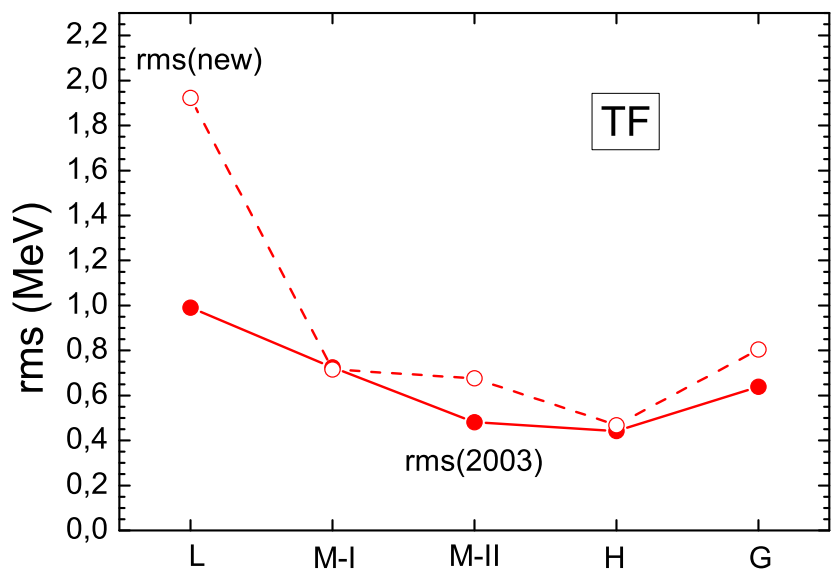

**Fig. 13**:

The HFB21 model (Fig. 14) shows a relatively good predictive power for the L, M-I and G regions and especially bad for the H region.

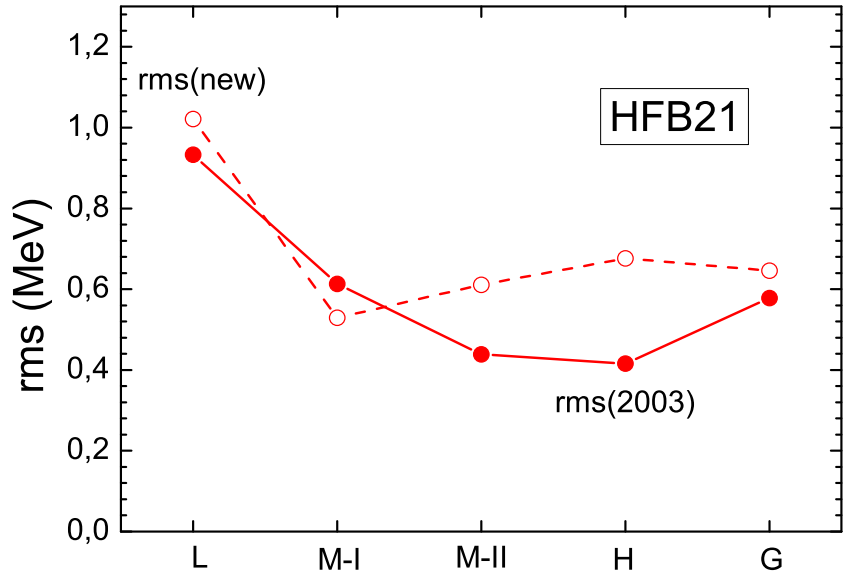

**Fig. 14**:

The GHFB model (Fig. 15) shows a good predictive power only in the M-II and G regions and especially bad in the L region.

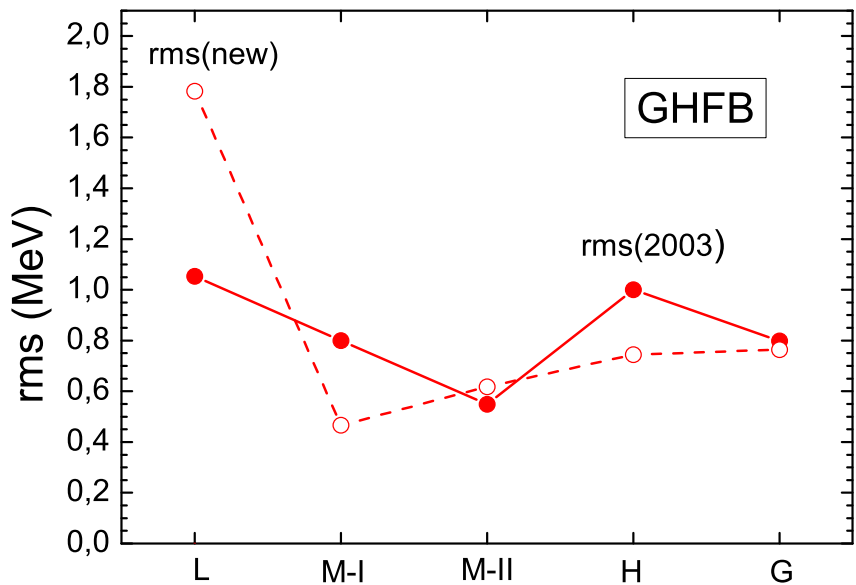

**Fig. 15**:

The DZ model (Fig. 16) shows a bad predictive power in all regions of nuclei and especially bad in the H region.

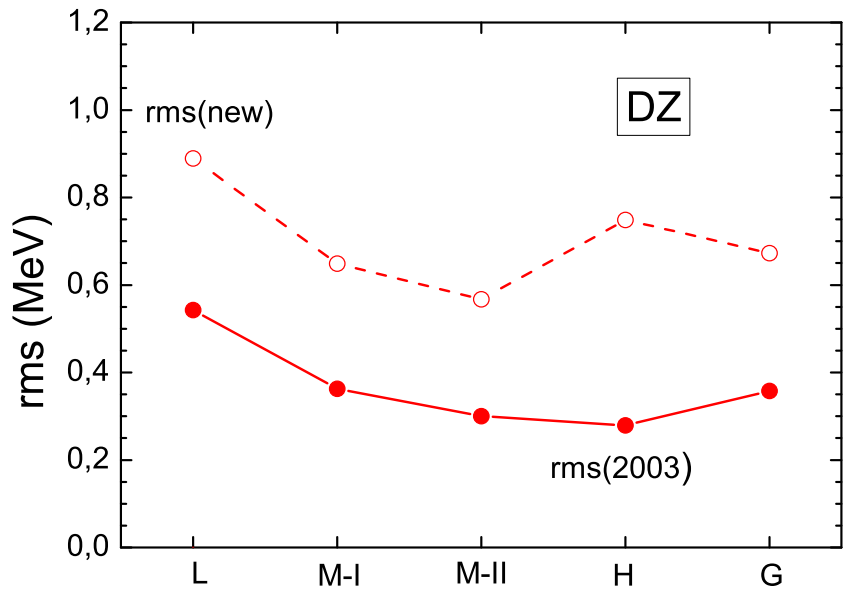

**Fig. 16**:

For the KTUY model (Fig. 17), a good predictive power is observed only in the M-II region and especially bad in the M-I region.

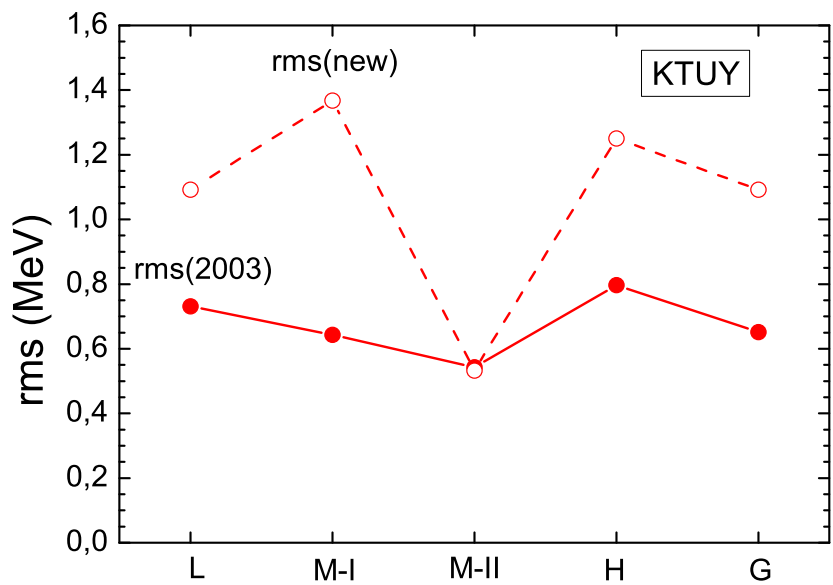

**Fig. 17**:

The INM model (Fig. 18) shows bad predictive power in all regions of nuclei and especially bad in the region M-I.

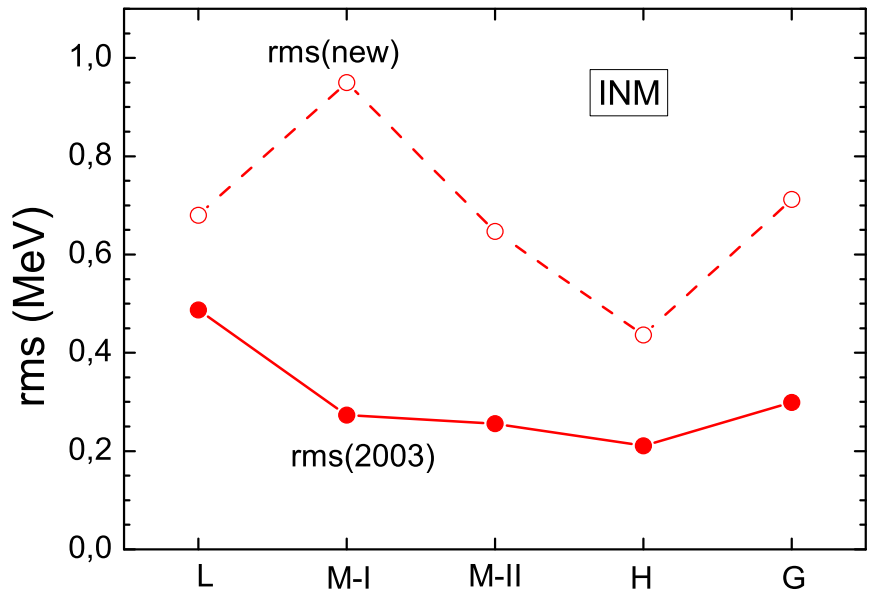

**Fig. 18**:

The WS3+ model (Fig. 19) shows rather bad predictive power in all regions of nuclei and especially bad in the L and M-I regions.

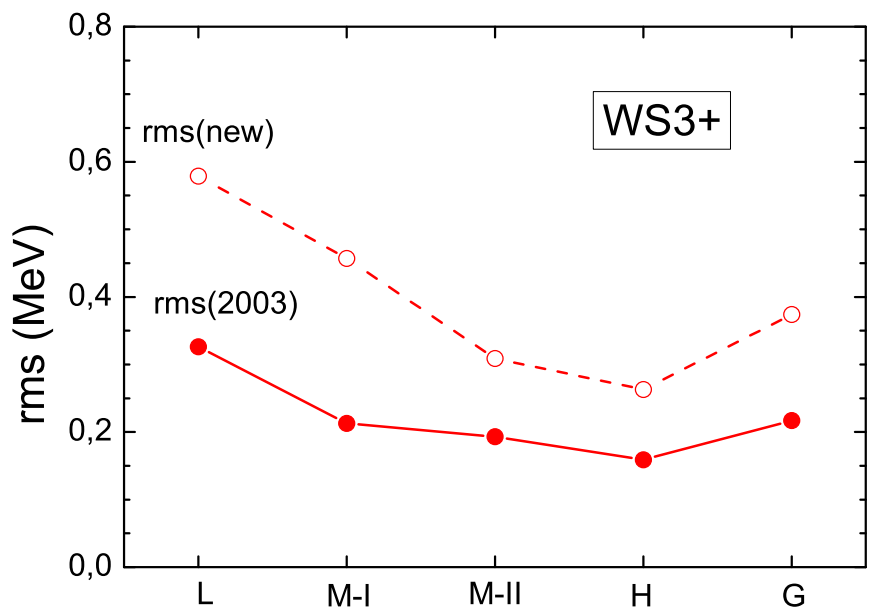

**Fig. 19**:

Finally, the WS4+ model (Fig. 20) shows a good predictive power in the M-II and H regions. This is quite different from the results obtained for the WS3+ model. This difference may look strange for these two models, which are similar to each other in many respects. It seems that the difference is, at least partly, due to the different data sets to which they were adjusted. The WS3+ model was fitted to masses of the older evaluation of 2003 [27] and the WS4+ to the recent one [26].

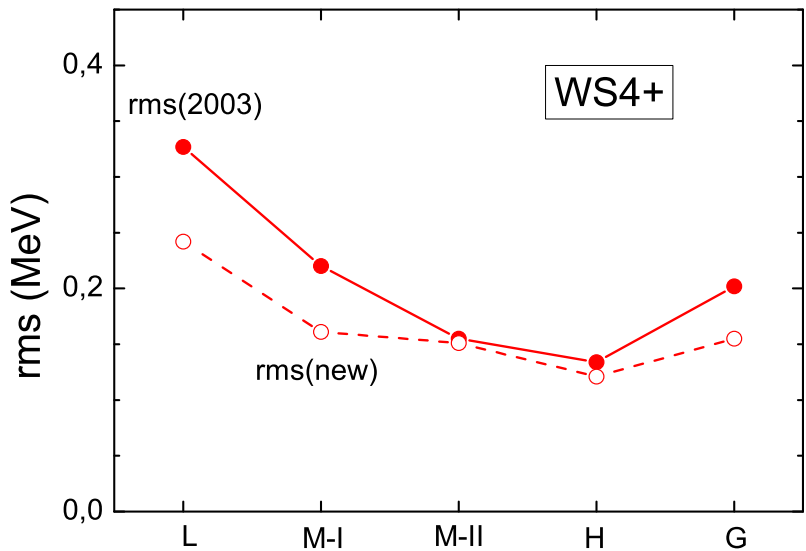

**Fig. 20**:

To conclude the two latter subsections, one may say the following: to have some reference point, one may assume that a good predictive power of a nuclear-mass model means that the description of unknown-yet-masses is of about the same accuracy as this for the masses already known. A test of this assumption by a set of ten various mass models, applied to five different regions of nuclei, shows that, generally, this is not the case. The definition is somewhat more often fulfilled by the macro-micro models than by the other ones. This is in line with a somewhat better description of already known masses by the macro-micro models and may be interpreted that the macroscopic part of the models (usually liquid drop) is physically reasonable.

## ACKNOWLEDGMENTS

We would like to thank the authors of the models discussed here for making the corresponding mass tables available to us. We are also grateful to Stephane Goriely, Jie Meng, Christoph Scheidenberger and Ning Wang for helpful discussions and correspondence. Support by the Polish-JINR(Dubna) Cooperation Programme, the GSI Helmholtzzentrum für Schwerionenforschung Gmbh, the European Science Foundation (within the EuroGenesis programme), the BMBF grant in the framework of the Internationale Zusammenarbeit in Bildung und Forschung (Projekt-Nr. 01DO12012), and the Helmholtz-CAS Joint Research Group (HCJRG-108) is gratefully acknowledged.

**Figures**

Figure captions:

Fig. 1. (Color online) Dependence of the rms discrepancy on the region of nuclei for the FRDM,TF, LSD, INM, WS3+, and WS4+ models. The symbols: L, M-I, M-II, H and G denote the regions of light, medium-I, medium-II, heavy and global nuclei, respectively.

Fig. 2. (Color online) Same as in Fig. 1, but for the GHFB, HFB21, KTUY, and DZ models.

Fig. 3. (Color online) The dependence of the average accuracy of the description of masses evaluated in 2012 [26], rms12, on a model in the medium-II region of nuclei. Each model is identified by the number (from 1 to 11) prescribed to it in Table A.

Fig. 4. (Color online) Same as in Fig. 3, but in the region of heavy nuclei.

Fig. 5. (Color online) Same as in Fig. 3, but in the region of the heaviest nuclei.

Fig. 6. (Color online) The dependence of the rms(2003) and rms(new) discrepancies on the model for the global region of nuclei. Each model is identified by the number (from 1 to 10) prescribed to it in Table B.

Fig. 7. (Color online) Same as in Fig. 6, but for the region of light nuclei.

FIG. 8. (Color online) Same as in Fig. 6, but for the medium-I region of nuclei.

Fig. 9. (Color online) Same as in Fig. 6, but for the medium-II region of nuclei.

Fig. 10. (Color online) Same as in Fig. 6, but for the region of heavy nuclei.

Fig. 11. (Color online) The dependence of the rms(2003) and rms(new) discrepancies on the region of nuclei for the LSD

model. The symbols: L, M-I, M-II, H and G denote the regions of light, medium-I, medium-II, heavy, and global nuclei, respectively.

Fig. 12. (Color online) Same as in Fig. 11, but for the FRDM model.

Fig. 13. (Color online) Same as in Fig. 11, but for the TF model.

Fig. 14. (Color online) Same as in Fig. 11, but for the HFB21 model.

Fig. 15. (Color online) Same as in Fig. 11, but for the GHFB model.

Fig. 16. (Color online) Same as in Fig. 11, but for the DZ model.

Fig. 17. (Color online) Same as in Fig. 11, but for the KTUY model.

Fig. 18. (Color online) Same as in Fig. 11, but for the INM model.

Fig. 19. (Color online) Same as in Fig. 11, but for the WS3+ model.

Fig. 20. (Color online) Same as in Fig. 11, but for the WS4+ model.

Fig. 21. (Color online) Map of the discrepancies between calculated and evaluated [26] masses in the region of light nuclei for the SLD model.

Fig. 22. (Color online) Same as in Fig. 21, but for the FRDM model.

Fig. 23. (Color online) Same as in Fig. 21, but for the FRDM12 model.

Fig. 24. (Color online) Same as in Fig. 21, but for the TF model.

Fig. 25. (Color online) Same as in Fig. 21, but for the HFB21 model.

Fig. 26. (Color online) Same as in Fig. 21, but for the GHFB model.

Fig. 27. (Color online) Same as in Fig. 21, but for the DZ model.

Fig. 28. (Color online) Same as in Fig. 21, but for the KTUY model.

Fig. 29. (Color online) Same as in Fig. 21, but for the INM model.

Fig. 30. (Color online) Same as in Fig. 21, but for the WS3+ model.

Fig. 31. (Color online) Same as in Fig. 21, but for the WS4+ model.

Fig. 32. (Color online) Map of the discrepancies between calculated and evaluated [26] masses in the region of medium-I nuclei for the SLD model.

Fig. 33. (Color online) Same as in Fig. 32, but for the FRDM model.

Fig. 34. (Color online) Same as in Fig. 32, but for the FRDM12 model.

Fig. 35. (Color online) Same as in Fig. 32, but for the TF model.

Fig. 36. (Color online) Same as in Fig. 32, but for the HFB21 model.

Fig. 37. (Color online) Same as in Fig. 32, but for the GHFB model.

Fig. 38. (Color online) Same as in Fig. 32, but for the DZ model.

Fig. 39. (Color online) Same as in Fig. 32, but for the KTUY model.

Fig. 40. (Color online) Same as in Fig. 32, but for the INM model.

Fig. 41. (Color online) Same as in Fig. 32, but for the WS3+ model.

Fig. 42. (Color online) Same as in Fig. 32, but for the WS4+ model.

Fig. 43. (Color online) Map of the discrepancies between calculated and evaluated [26] masses in the region of medium-II nuclei for the SLD model.

Fig. 44. (Color online) Same as in Fig. 43, but for the FRDM model.

Fig. 45. (Color online) Same as in Fig. 43, but for the FRDM12 model.

Fig. 46. (Color online) Same as in Fig. 43, but for the TF model.

Fig. 47. (Color online) Same as in Fig. 43, but for the HFB21 model.

Fig. 48. (Color online) Same as in Fig. 43, but for the GHFB model.

Fig. 49. (Color online) Same as in Fig. 43, but for the DZ model.

Fig. 50. (Color online) Same as in Fig. 43, but for the KTUY model.

Fig. 51. (Color online) Same as in Fig. 43, but for the INM model.

Fig. 52. (Color online) Same as in Fig. 43, but for the WS3+ model.

Fig. 53. (Color online) Same as in Fig. 43, but for the WS4+ model.

Fig. 54. (Color online) Map of the discrepancies between calculated and evaluated [26] masses in the region of heavy nuclei for the SLD model.

Fig. 55. (Color online) Same as in Fig. 54, but for the FRDM model.

Fig. 56. (Color online) Same as in Fig. 54, but for the FRDM12 model.

Fig. 57. (Color online) Same as in Fig. 54, but for the TF model.

Fig. 58. (Color online) Same as in Fig. 54, but for the HFB21 model.

Fig. 59. (Color online) Same as in Fig. 54, but for the GHFB model.

Fig. 60. (Color online) Same as in Fig. 54, but for the DZ model.

Fig. 61. (Color online) Same as in Fig. 54, but for the KTUY model.

Fig. 62. (Color online) Same as in Fig. 54, but for the INM model.

Fig. 63. (Color online) Same as in Fig. 54, but for the WS3+ model.

Fig. 64. (Color online) Same as in Fig. 54, but for the WS4+ model.

Fig. 65. (Color online) Same as in Fig. 54, but for the HN model and decreased region of heavy nuclei.

## 5. Maps of the accuracy of the description of nuclear mass by various models and in various regions of nuclear chart

*5.1. Light nuclei*

For light nuclei, the accuracy obtained with the use of eleven global models is shown in Table A and illustrated in the figures from Fig. 21 to Fig. 31. The first orientation obtained from the rms of the discrepancies given in Table A and Figs. 1 and 2, one may expect that the worst accuracy may be observed in this region for the models FRDM, TF, LSD and GHFB. Really, figures 21 to 24 show that the masses obtained by these models may be by about 3 MeV smaller or by about 4 MeV larger than the measured ones. Additionally, the positions of the nuclei with too small and too large calculated masses may be quite close to each other. In other words, the discrepancy changes quite fast with the change of $Z$ and $N$.

On the other side, it is seen in Figs. 30 and 31 that the discrepancies are much smaller and change much less with the changes of $Z$ and $N$ in the case of the WS3+ and the WS4+ models.

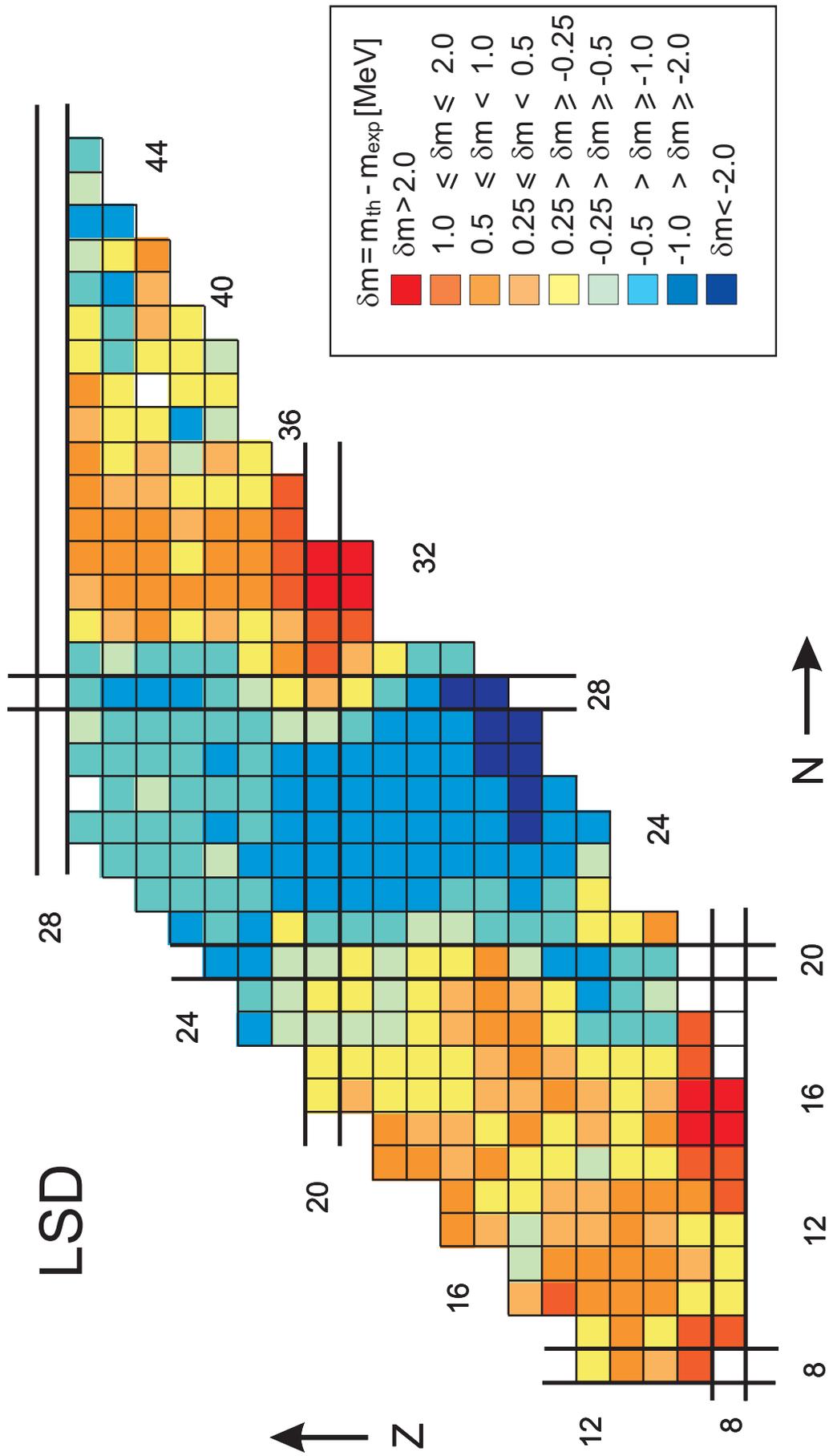

**Fig. 21:**

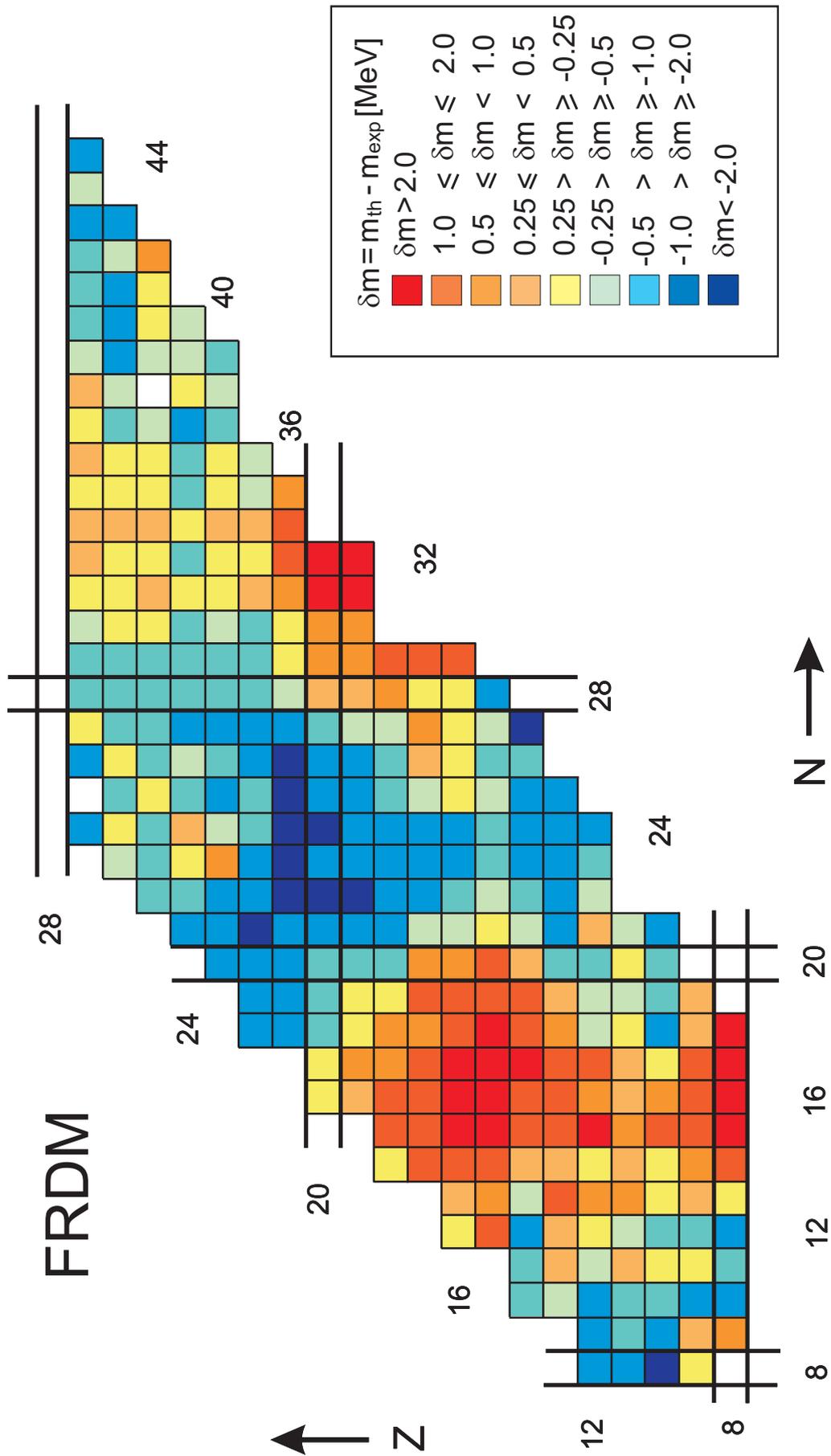

Fig. 22:

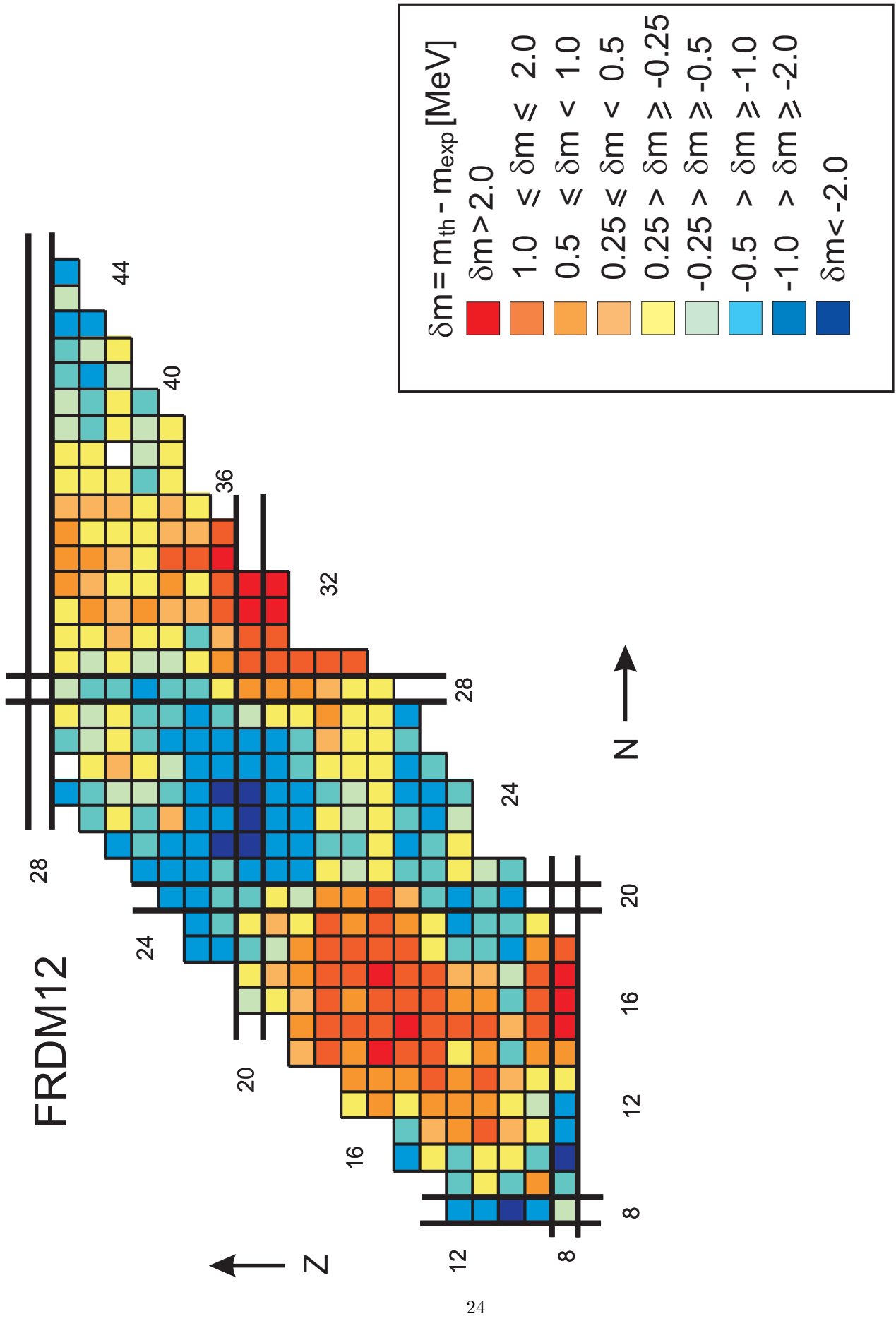

Fig. 23:

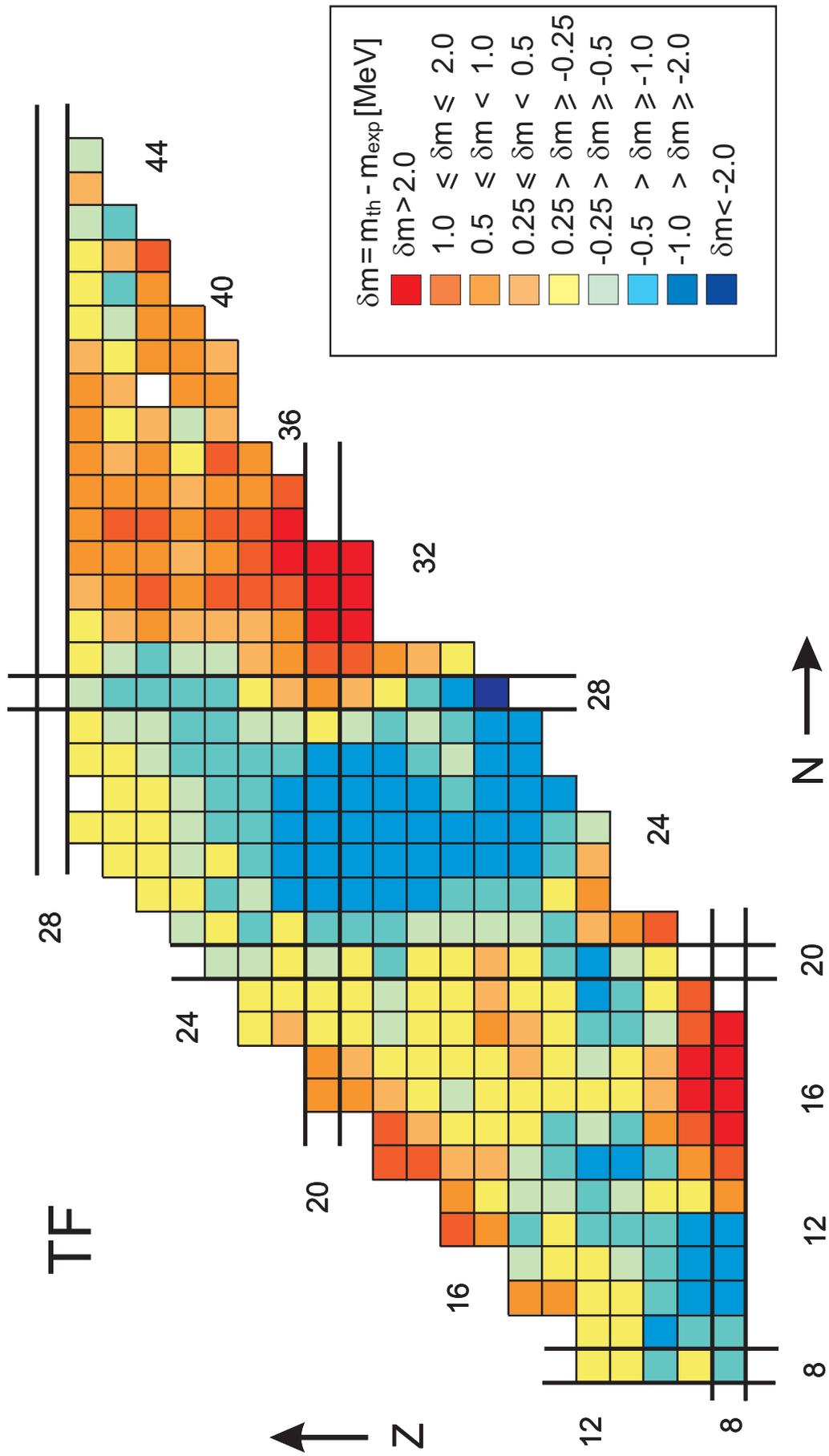

Fig. 24:

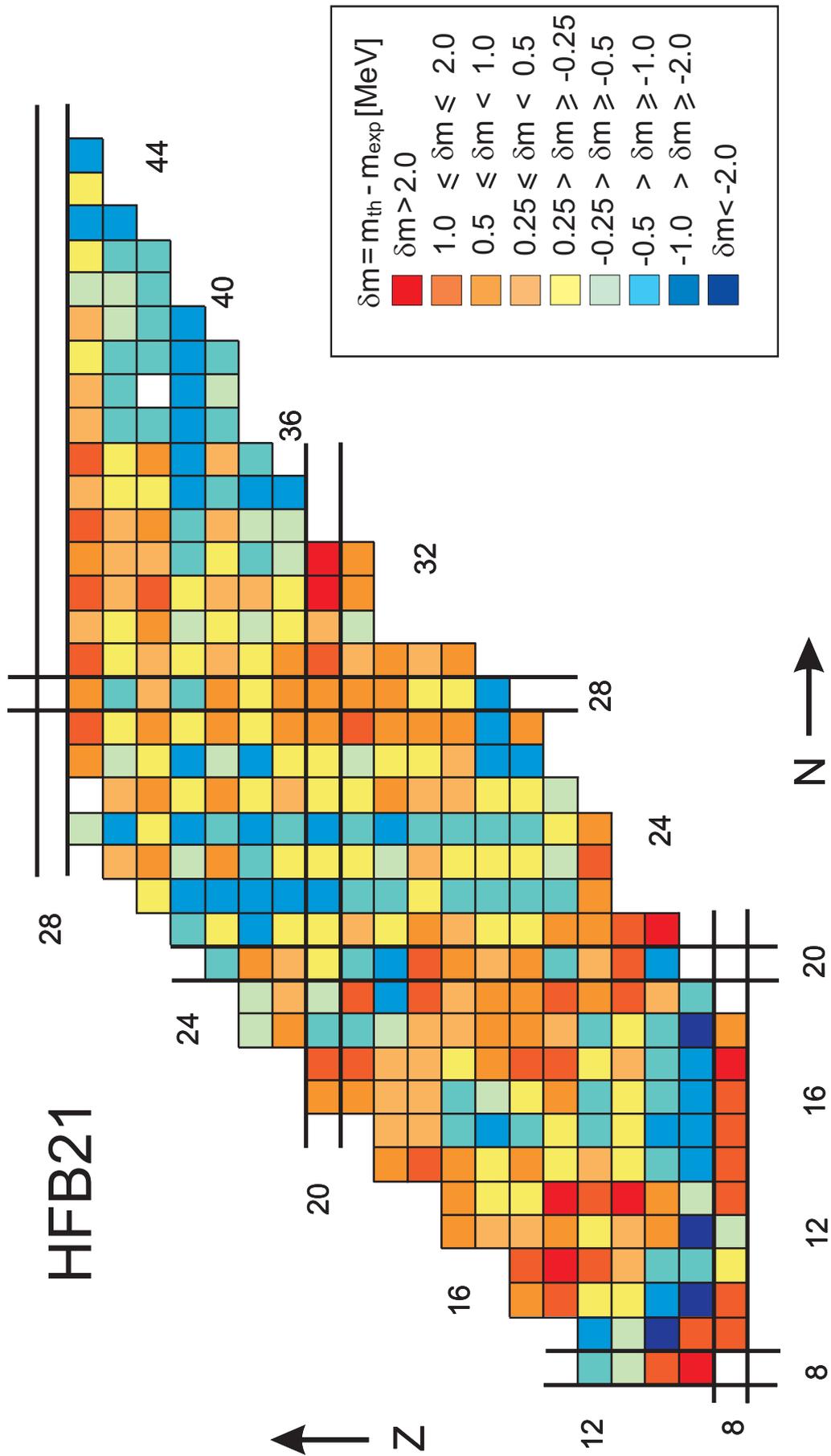

Fig. 25:

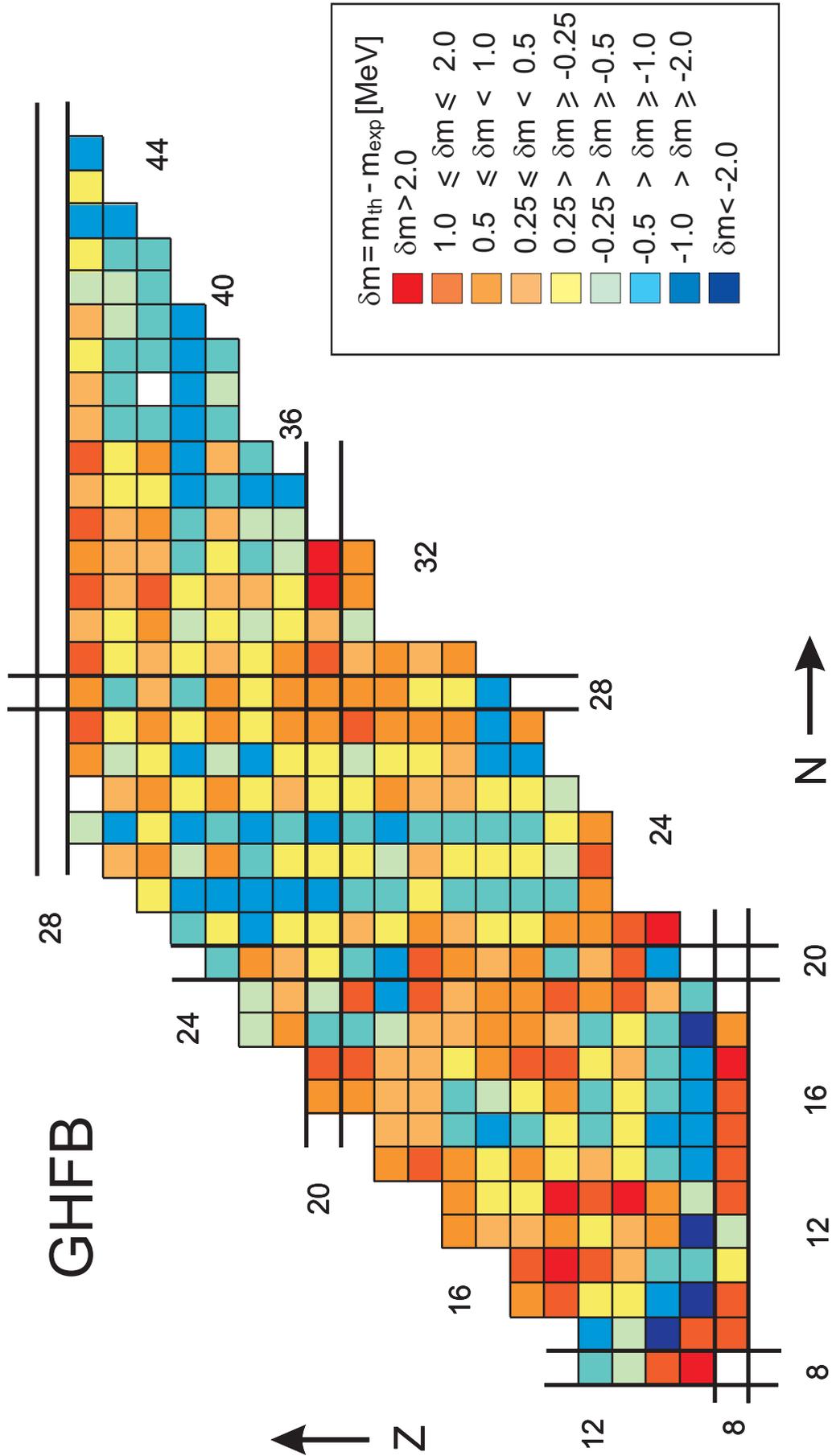

Fig. 26:

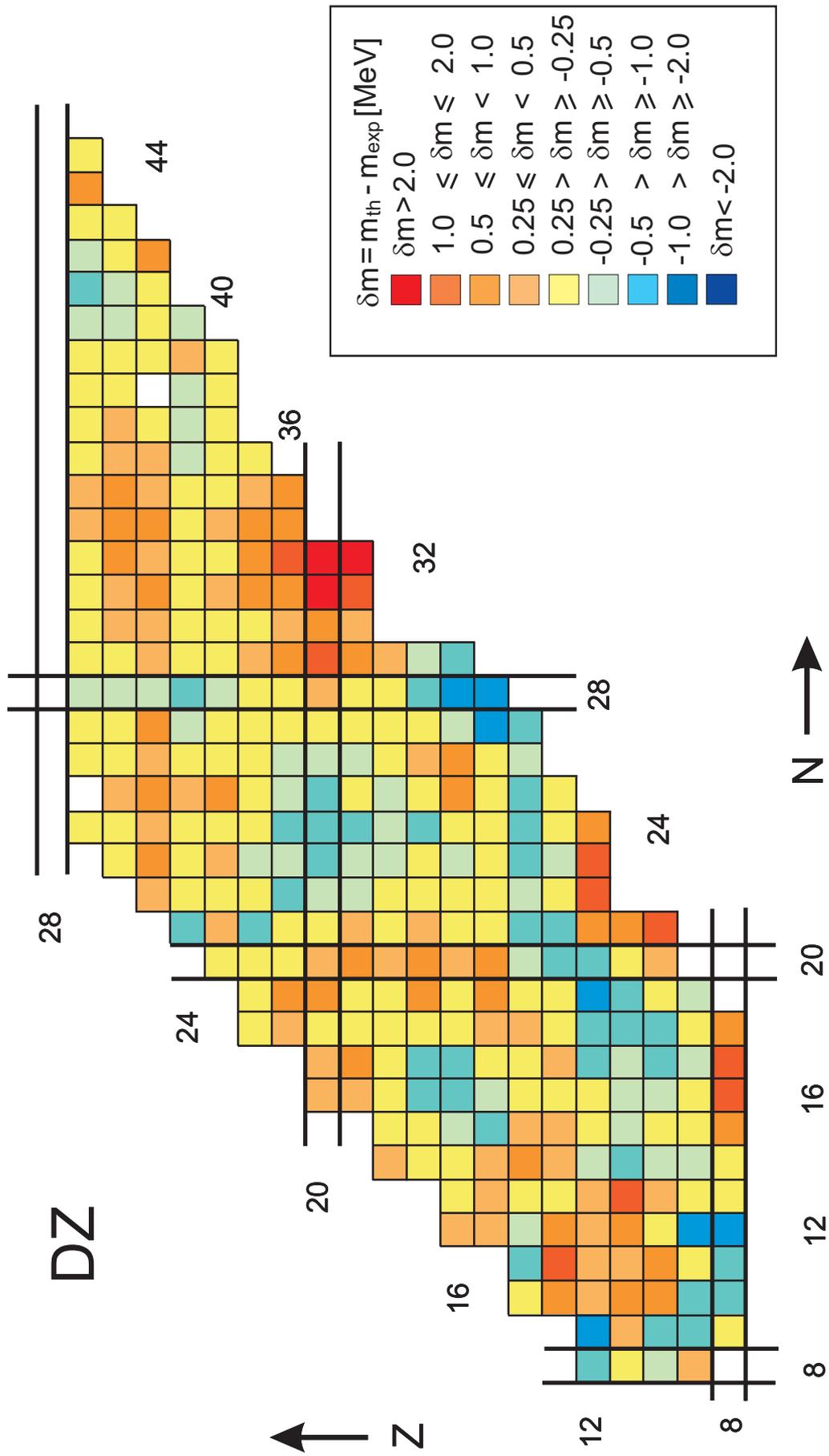

Fig. 27:

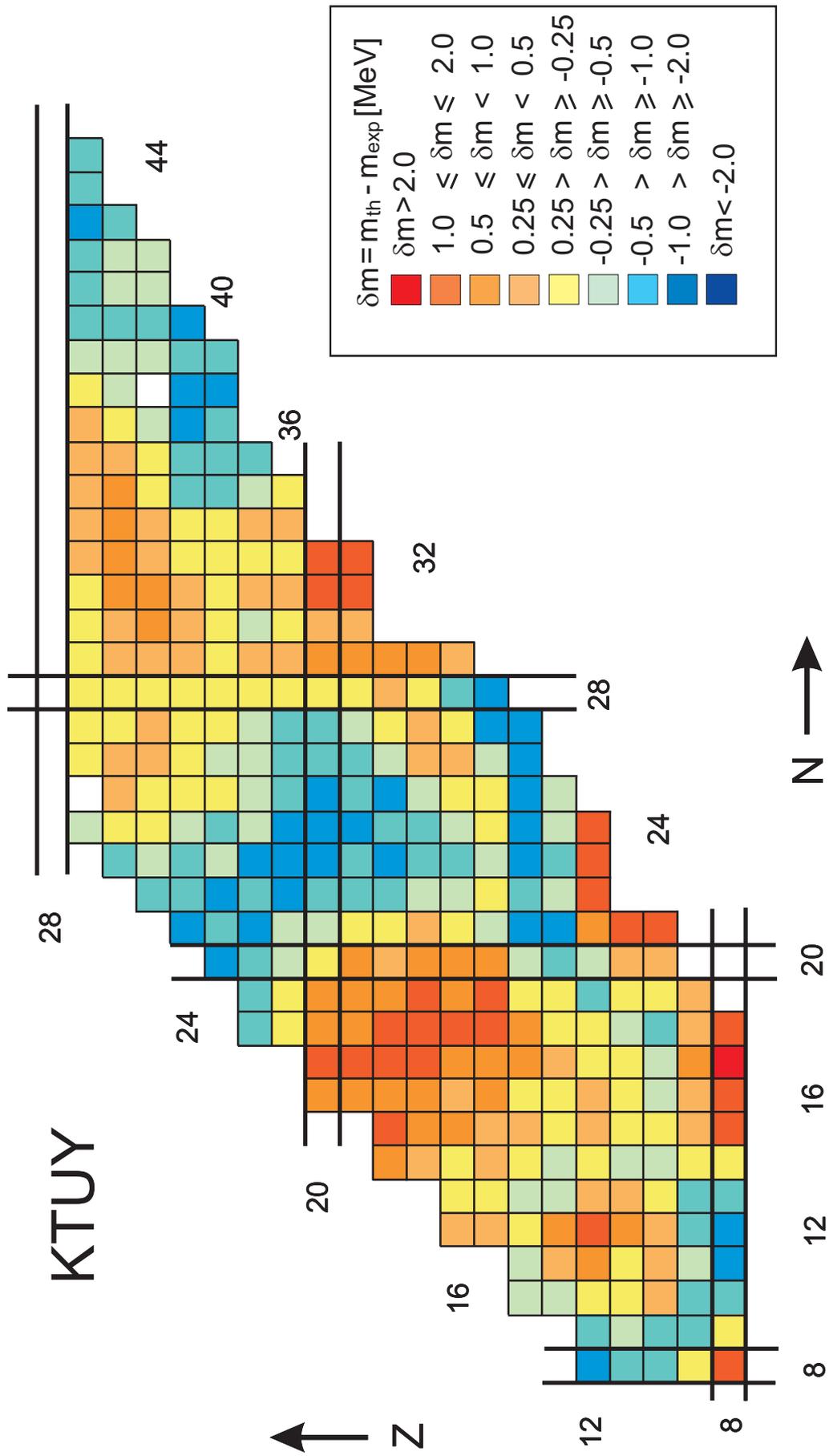

Fig. 28:

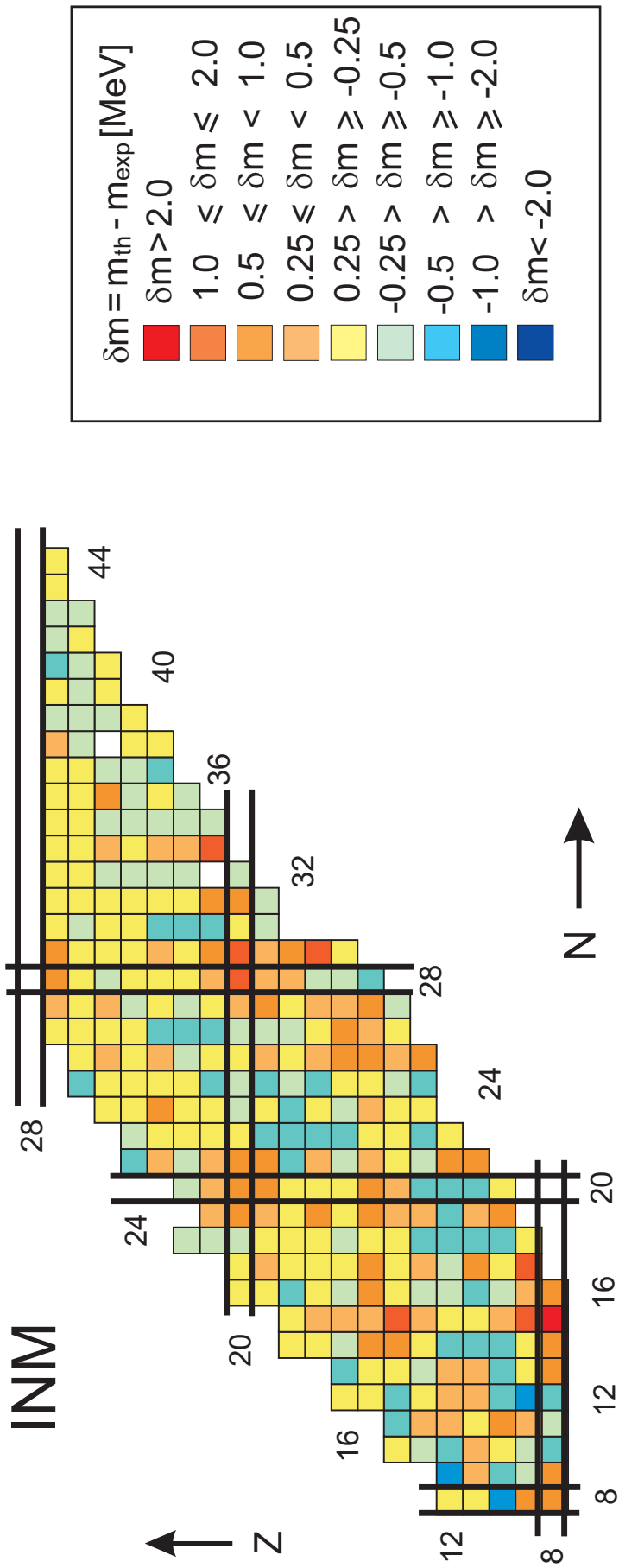

Fig. 29:

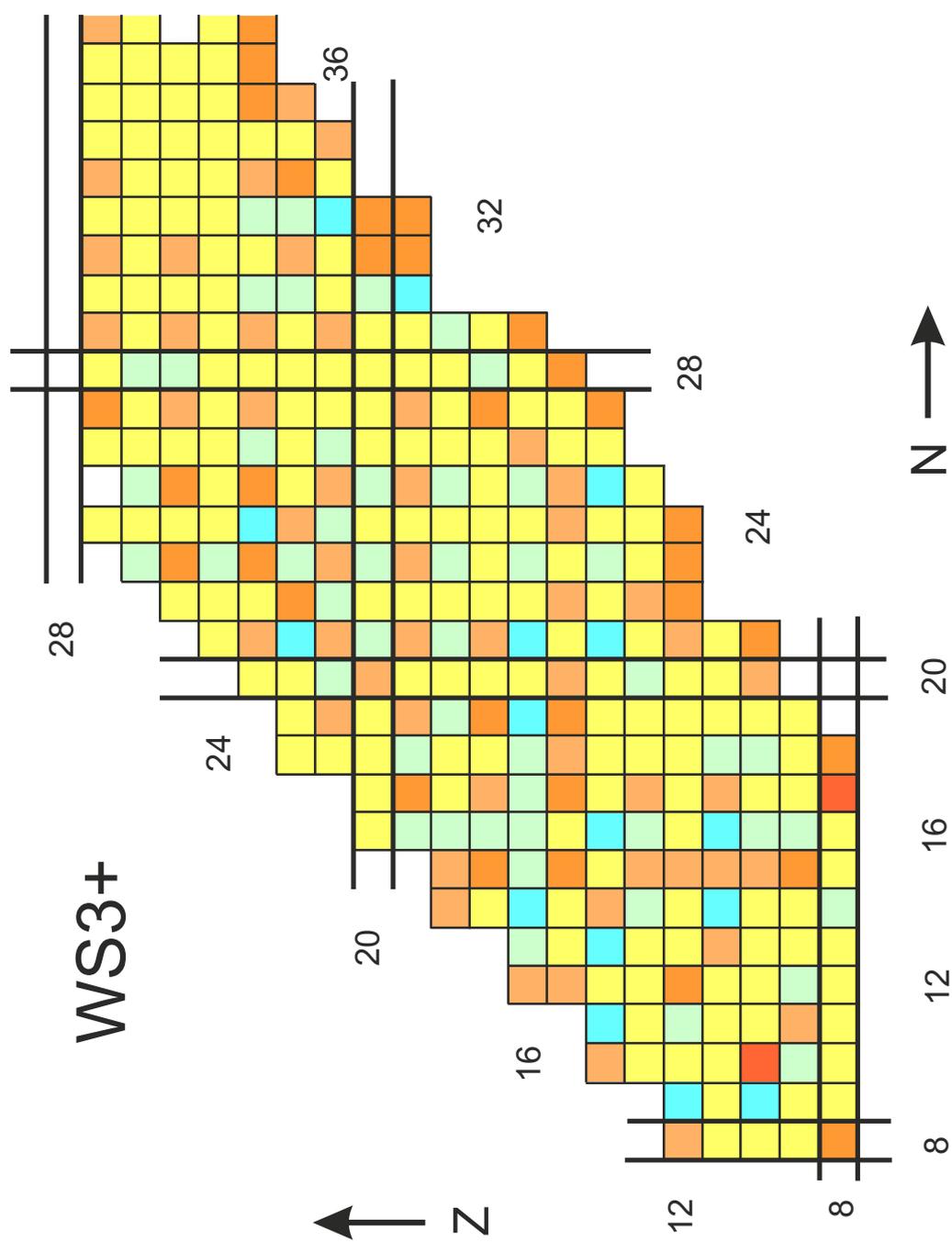

Fig. 30:

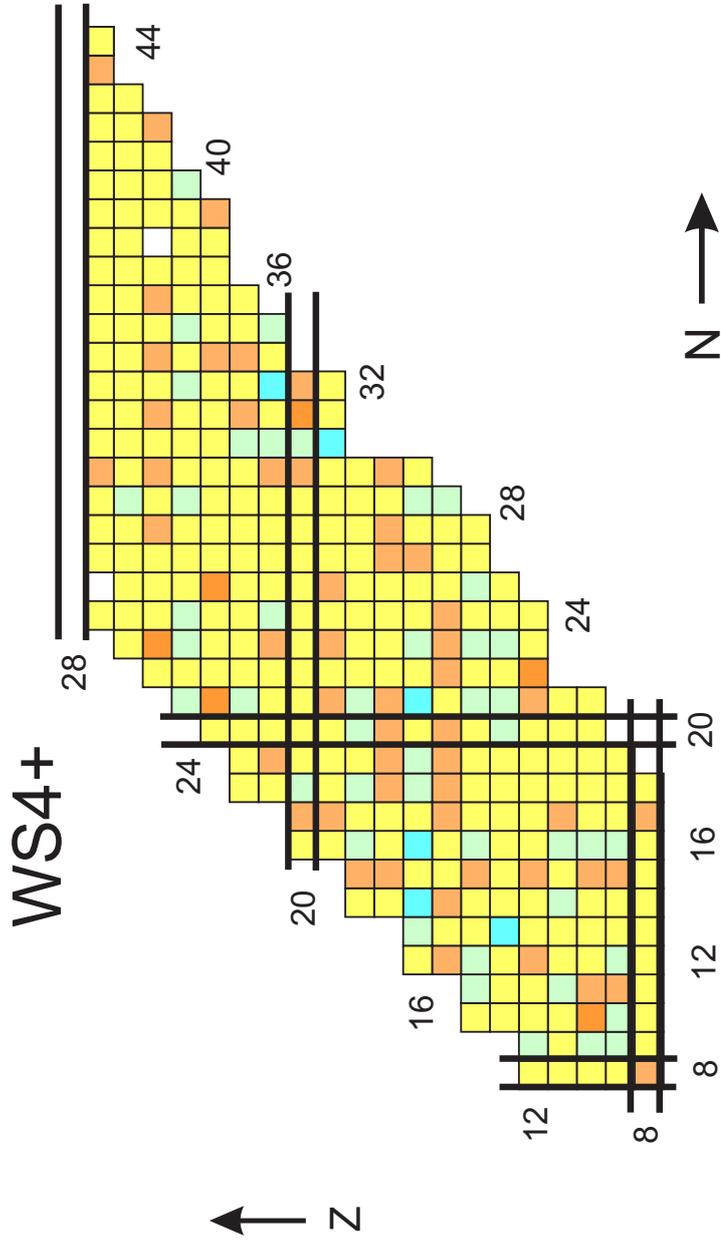

Fig. 31:

*5.2. Medium-I nuclei*

According to Table A and Figs. 1 and 2 masses of these nuclei may be expected to be described with a better accuracy than the light nuclei. Really, this is clearly seen when comparing the respective figures. For example, it is seen in Fig. 32 that the discrepancies are much smaller and they change much slower with the changes of $Z$ and $N$ than in Fig. 21. Similar, although in a smaller degree, is seen when comparing Fig. 42 with Fig. 31.

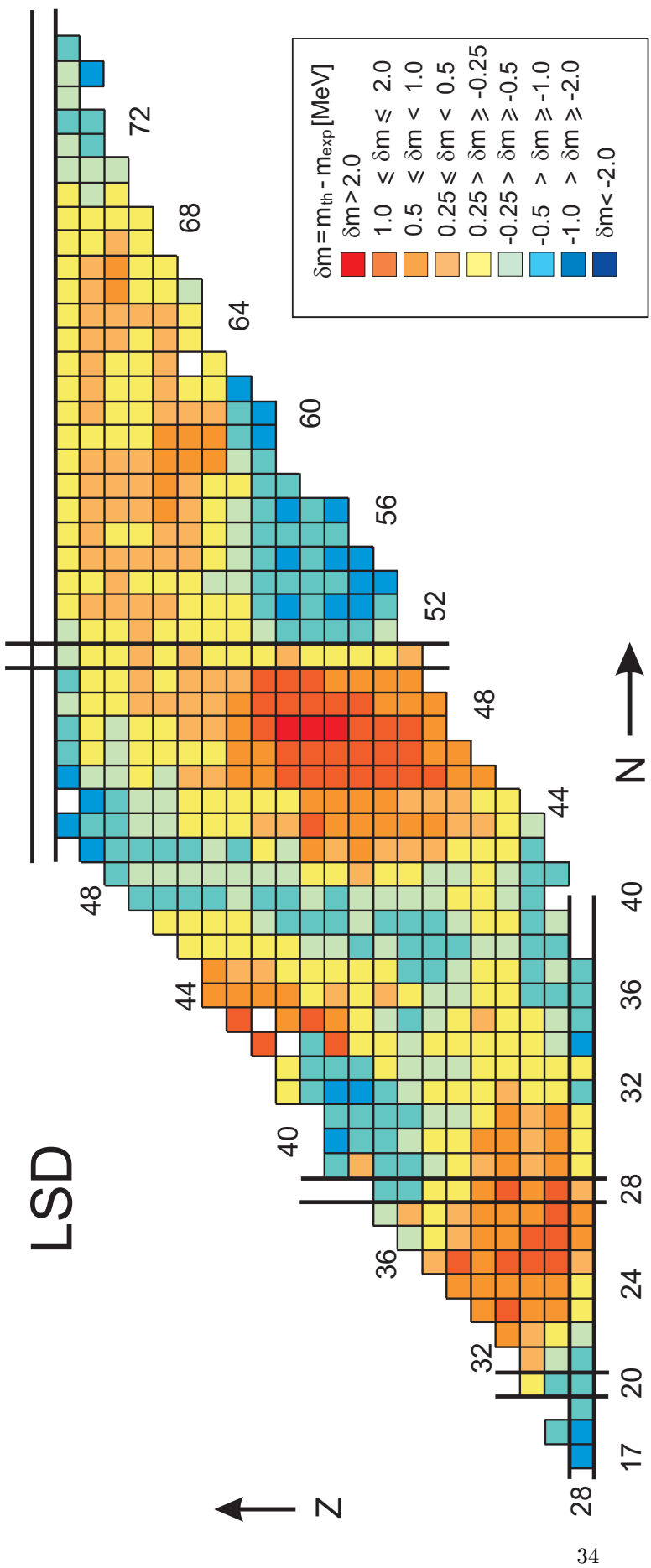

Fig. 32:

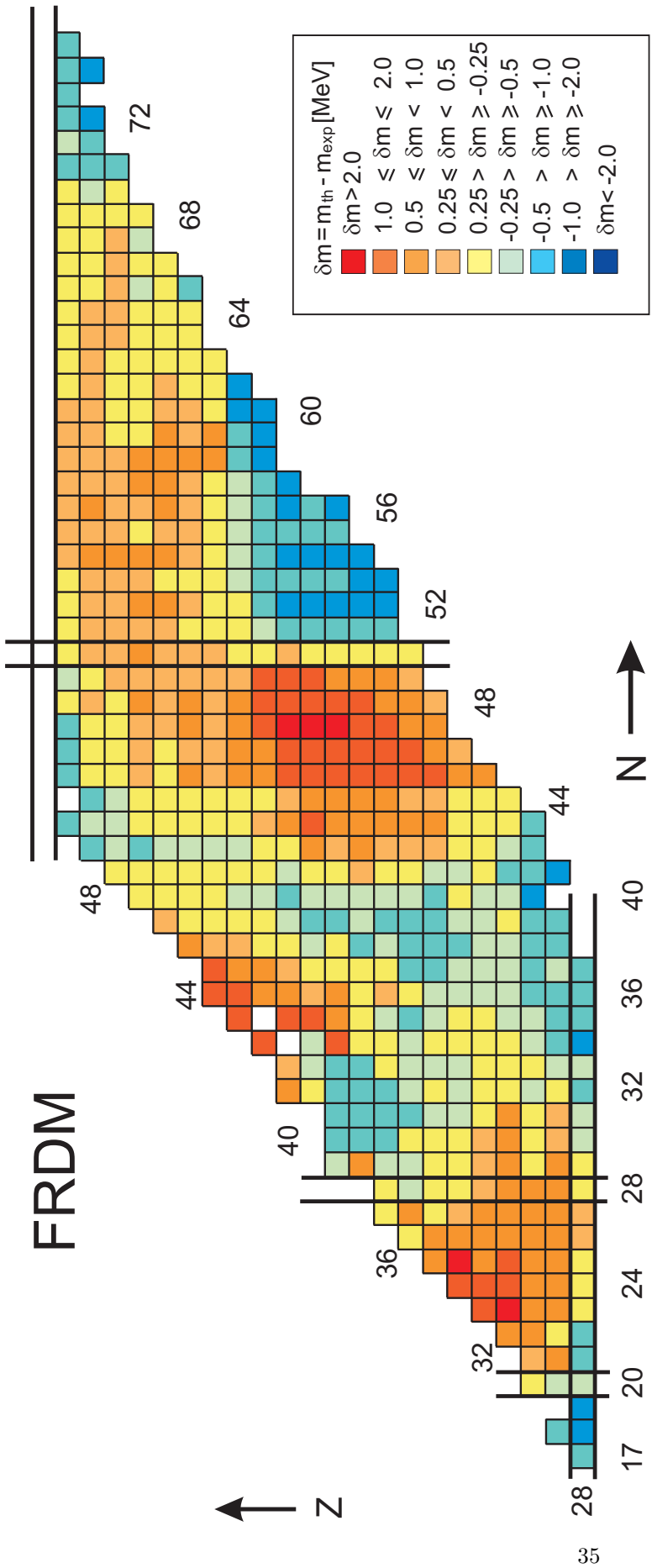

**Fig. 33:**

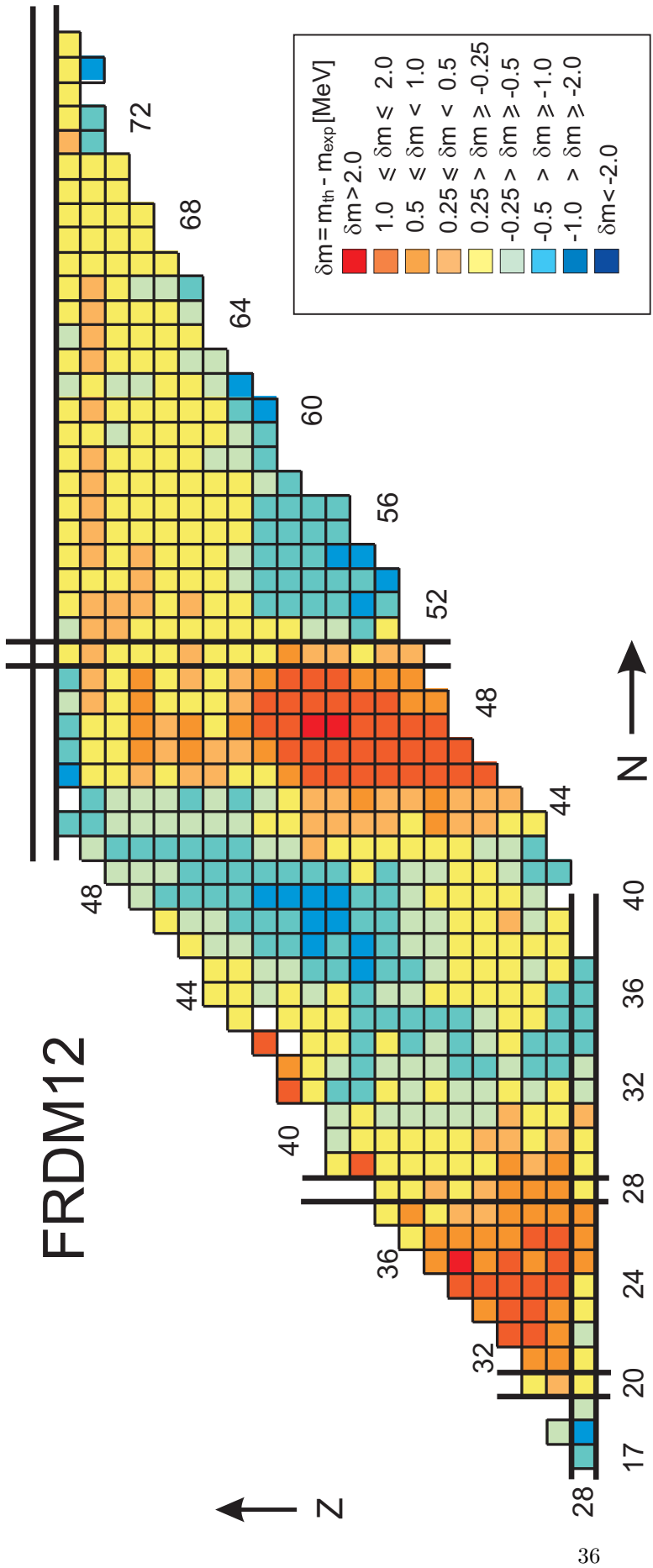

**Fig. 34:**

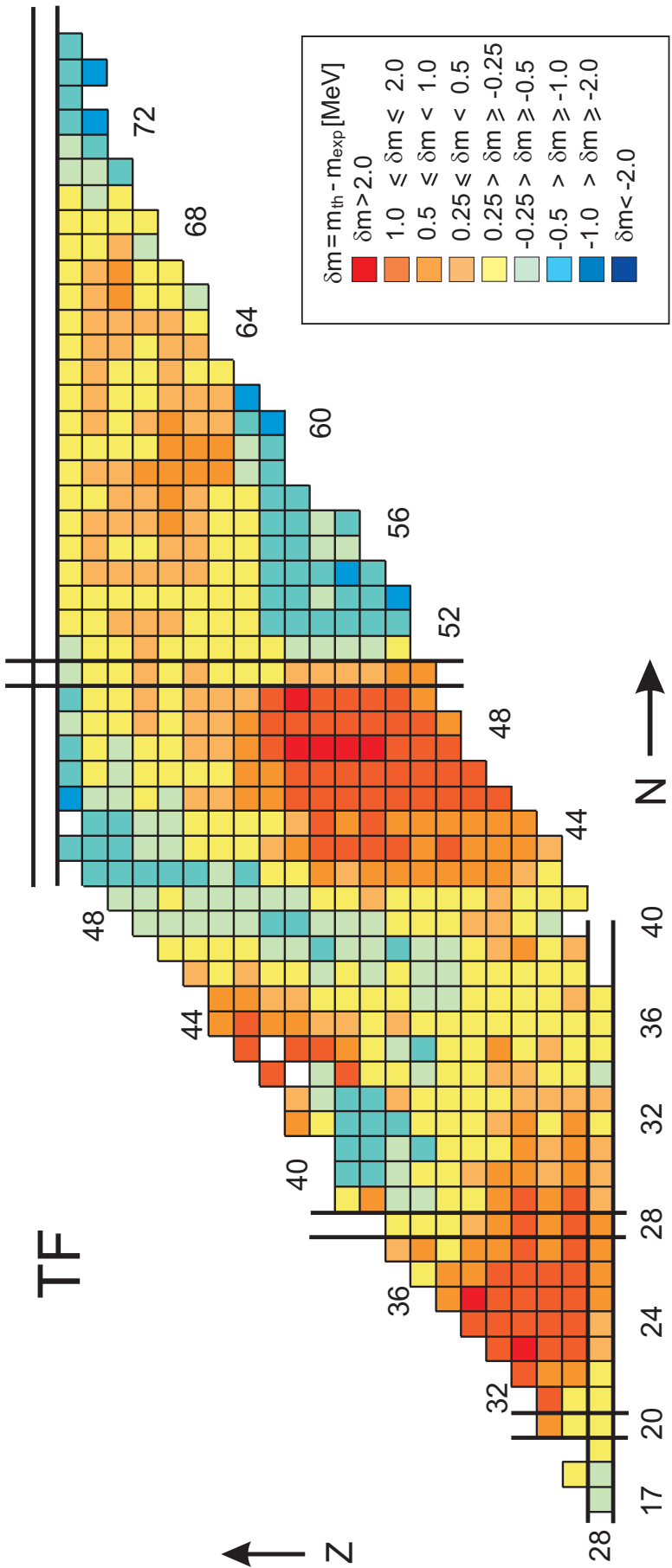

**Fig. 35:**

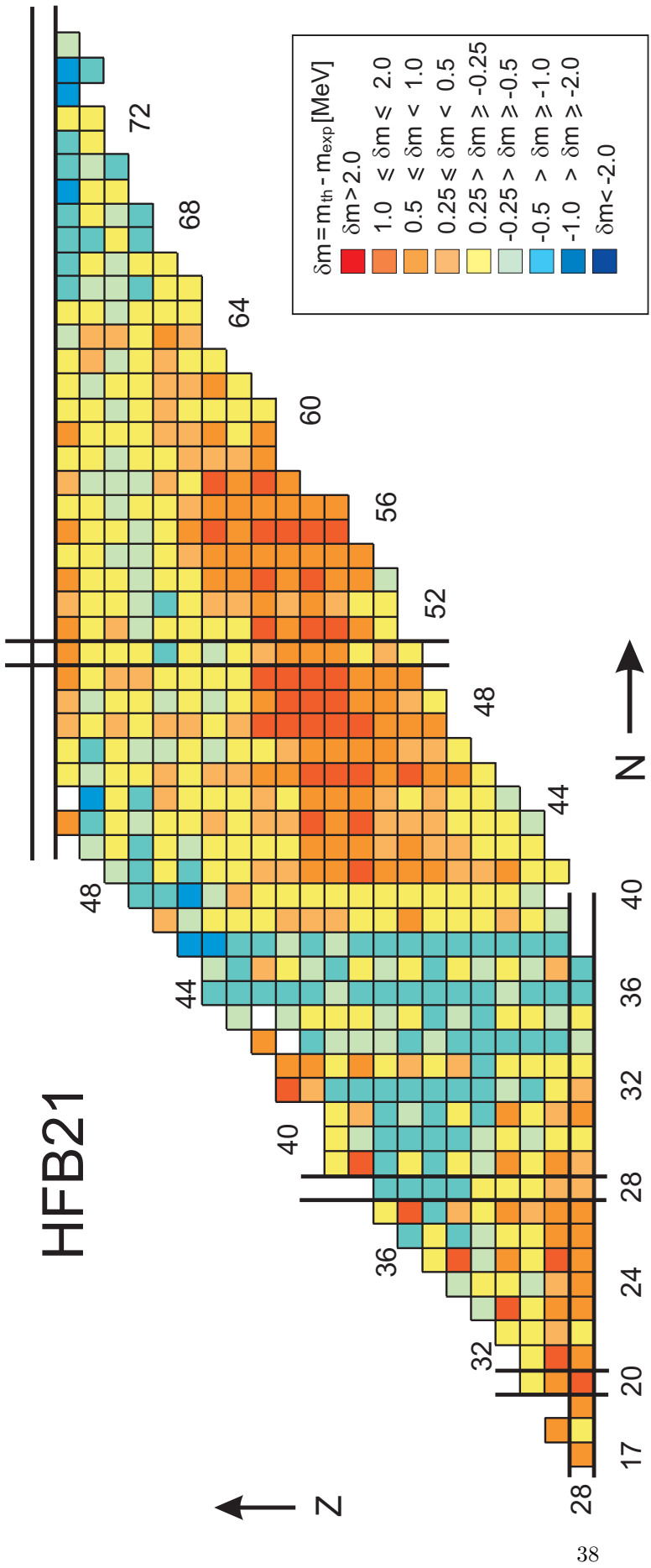

Fig. 36:

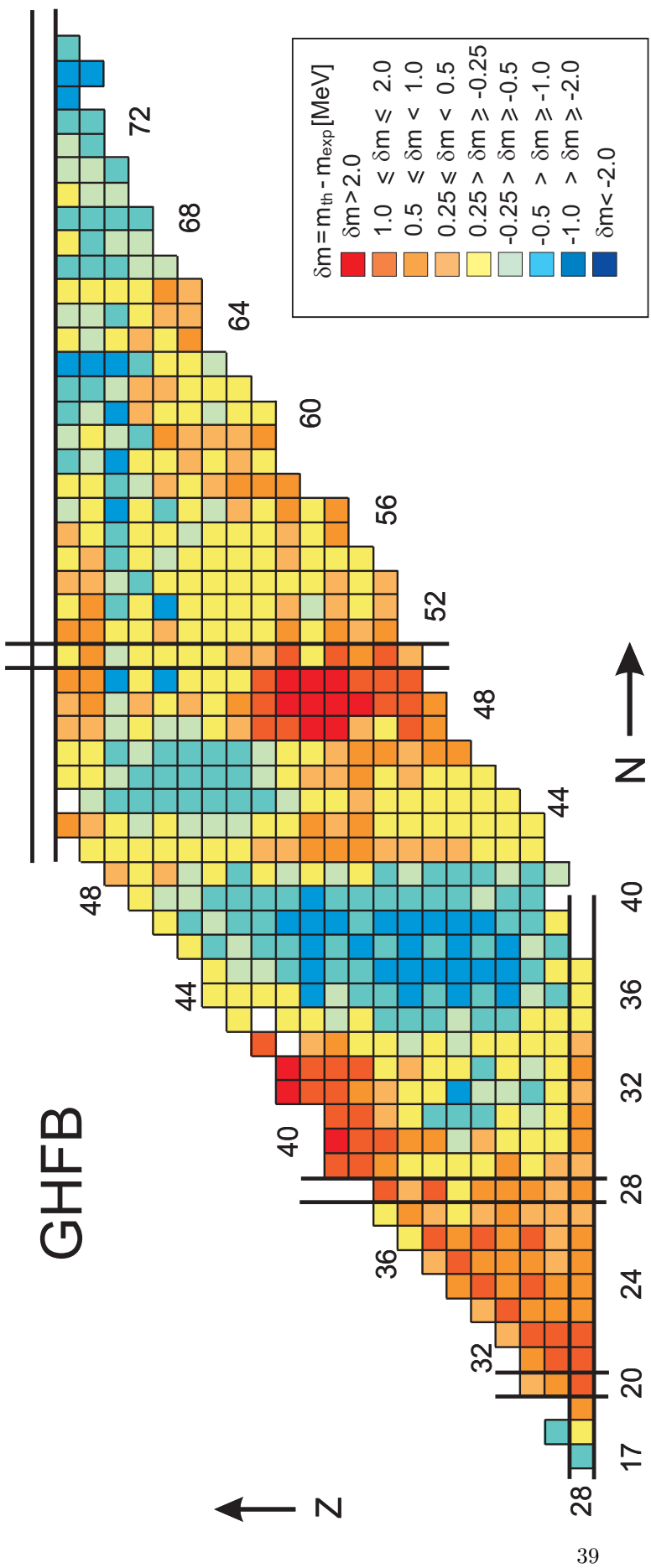

Fig. 37:

Fig. 38:

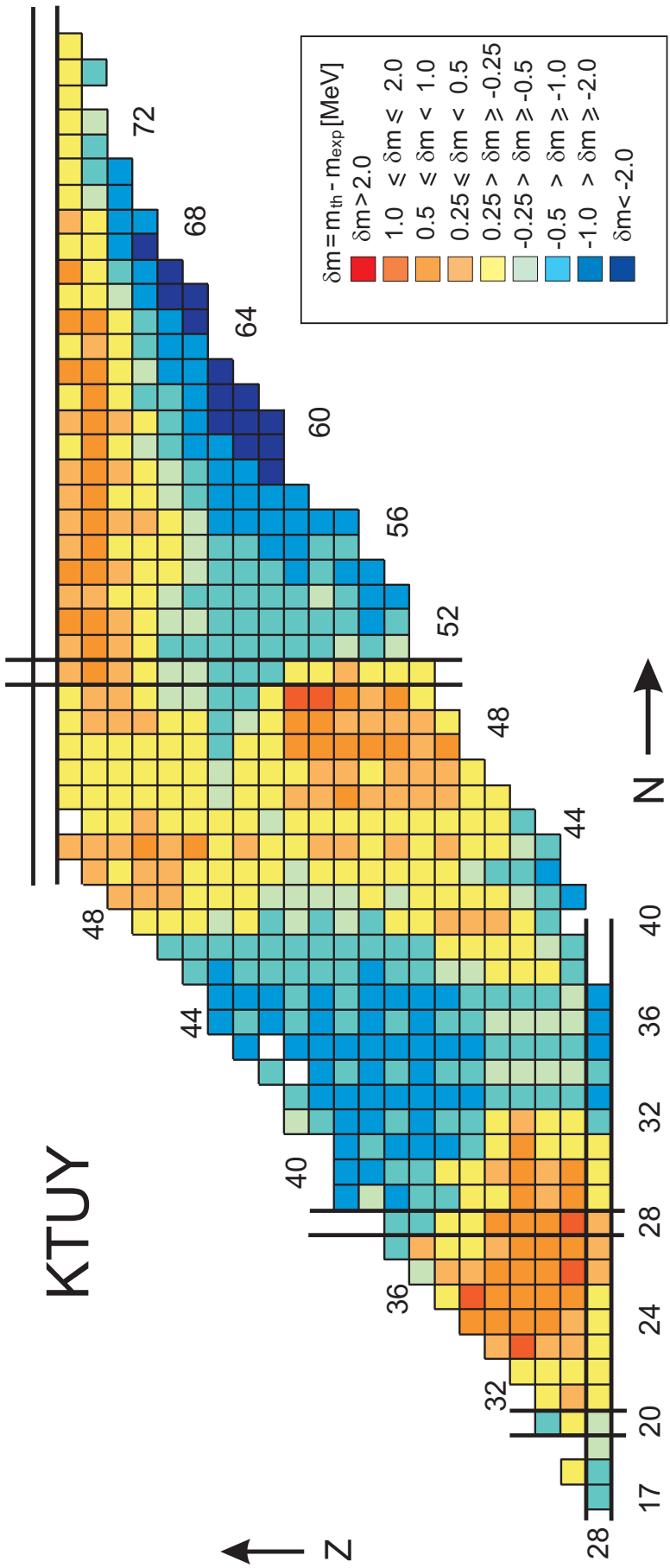

Fig. 39:

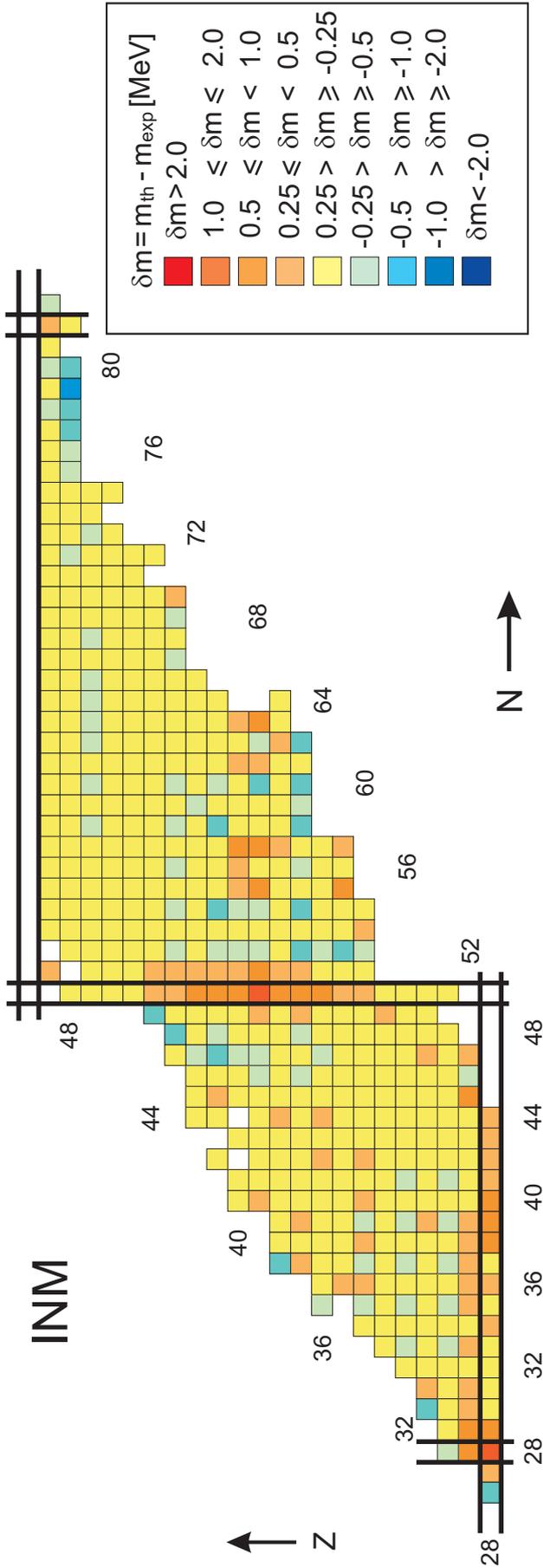

**Fig. 40:**

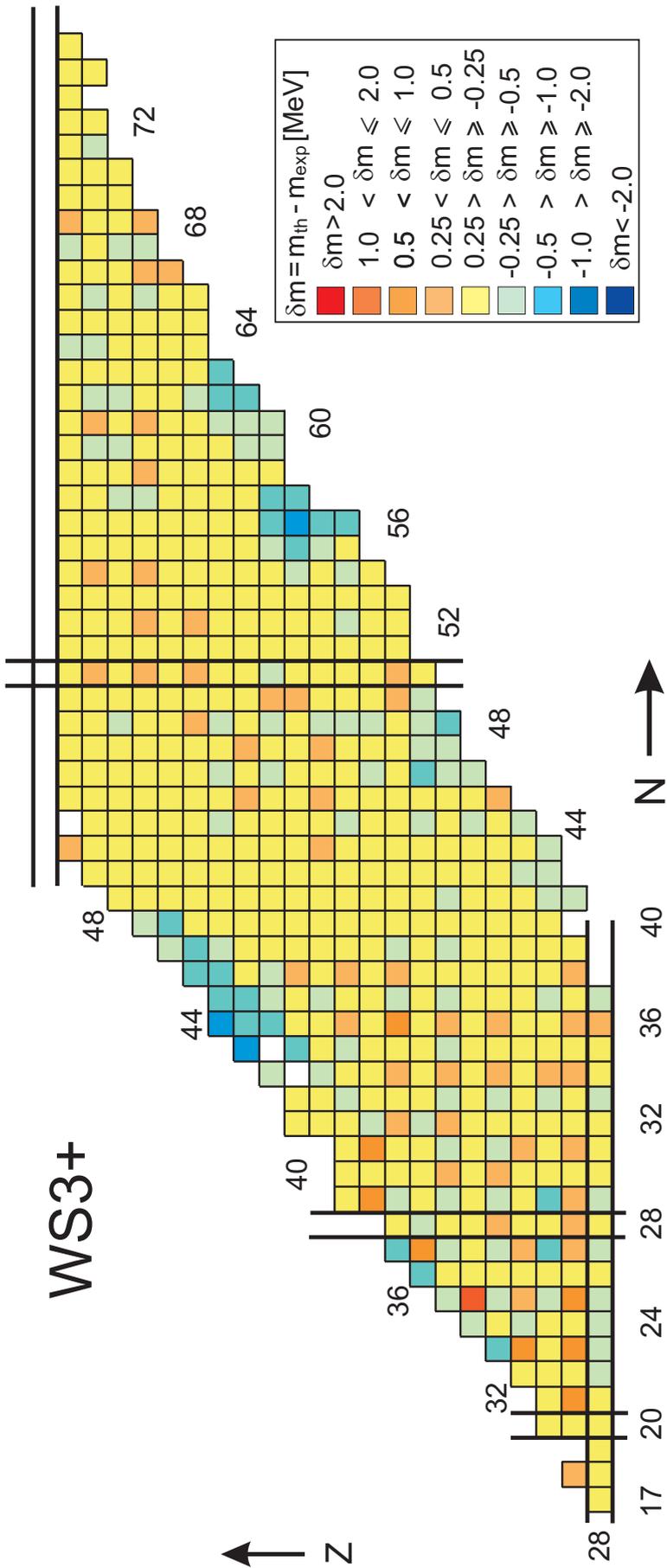

Fig. 41:

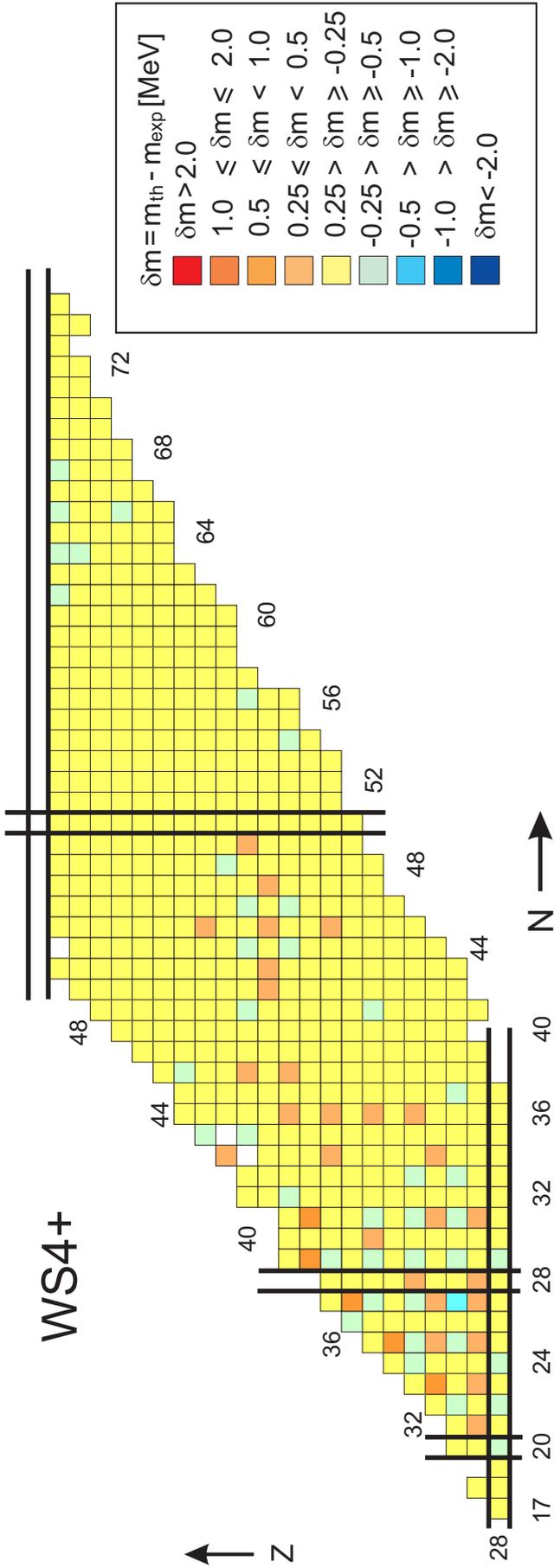

**Fig. 42:**

*5.3. Medium-II nuclei*

The tendency of the increasing the accuracy of the description of nuclear masses with increasing mass of nuclei is again seen when comparing figures showing the accuracy obtained in this region with the corresponding ones obtained for the mediun-I nuclei.

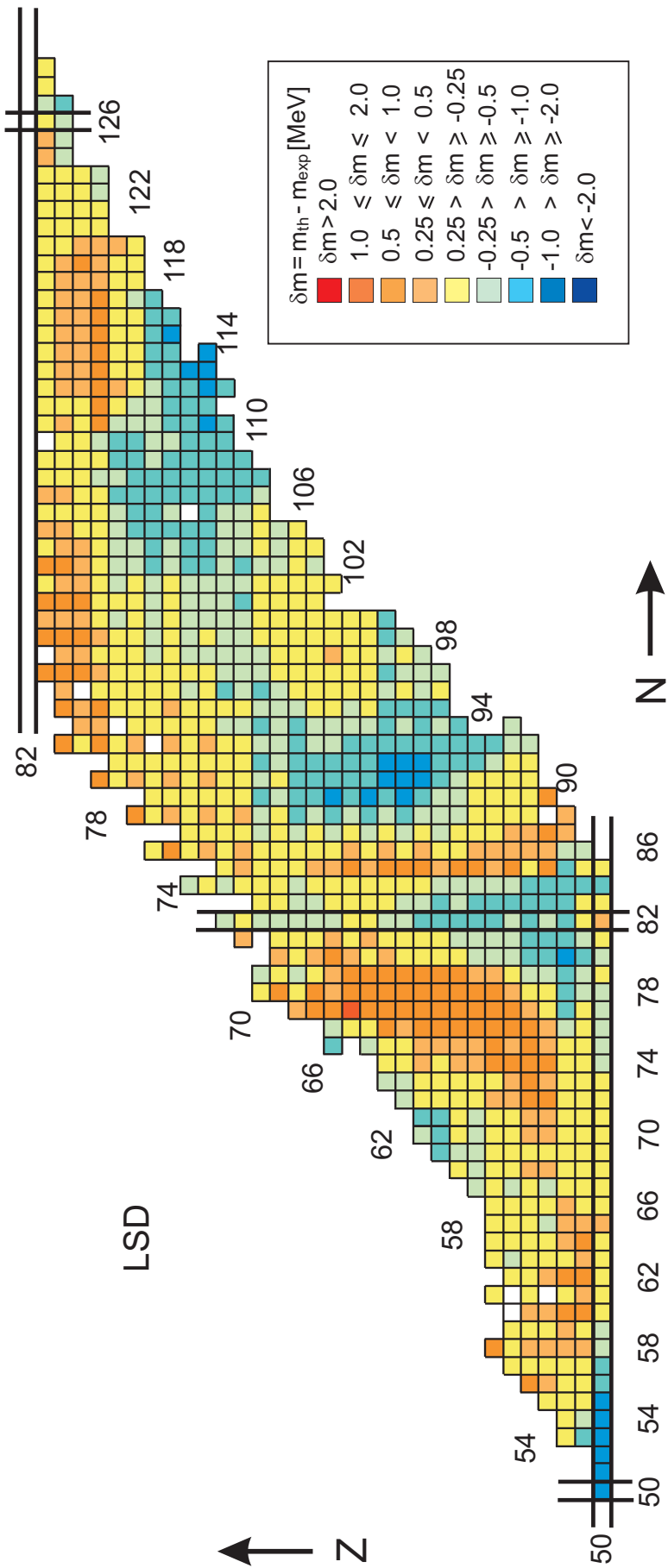

Fig. 43:

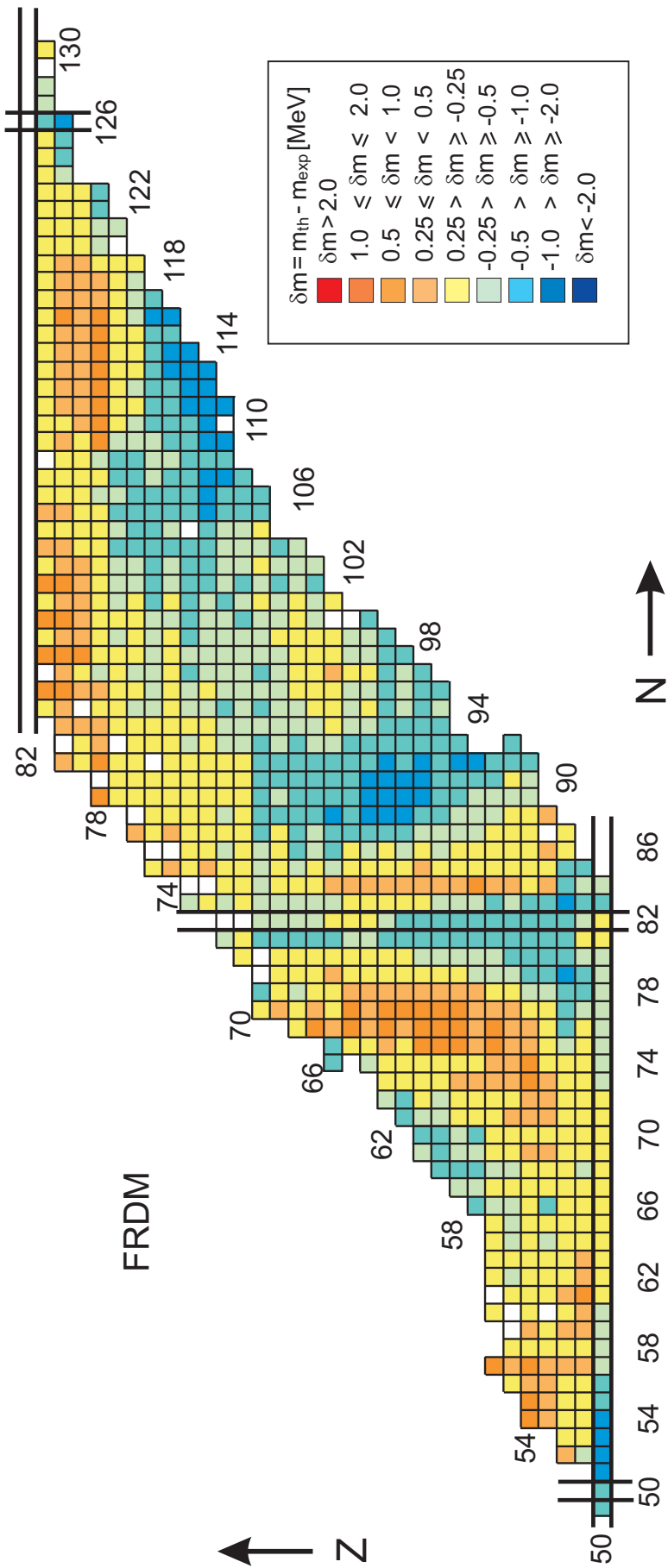

**Fig. 44:**
47

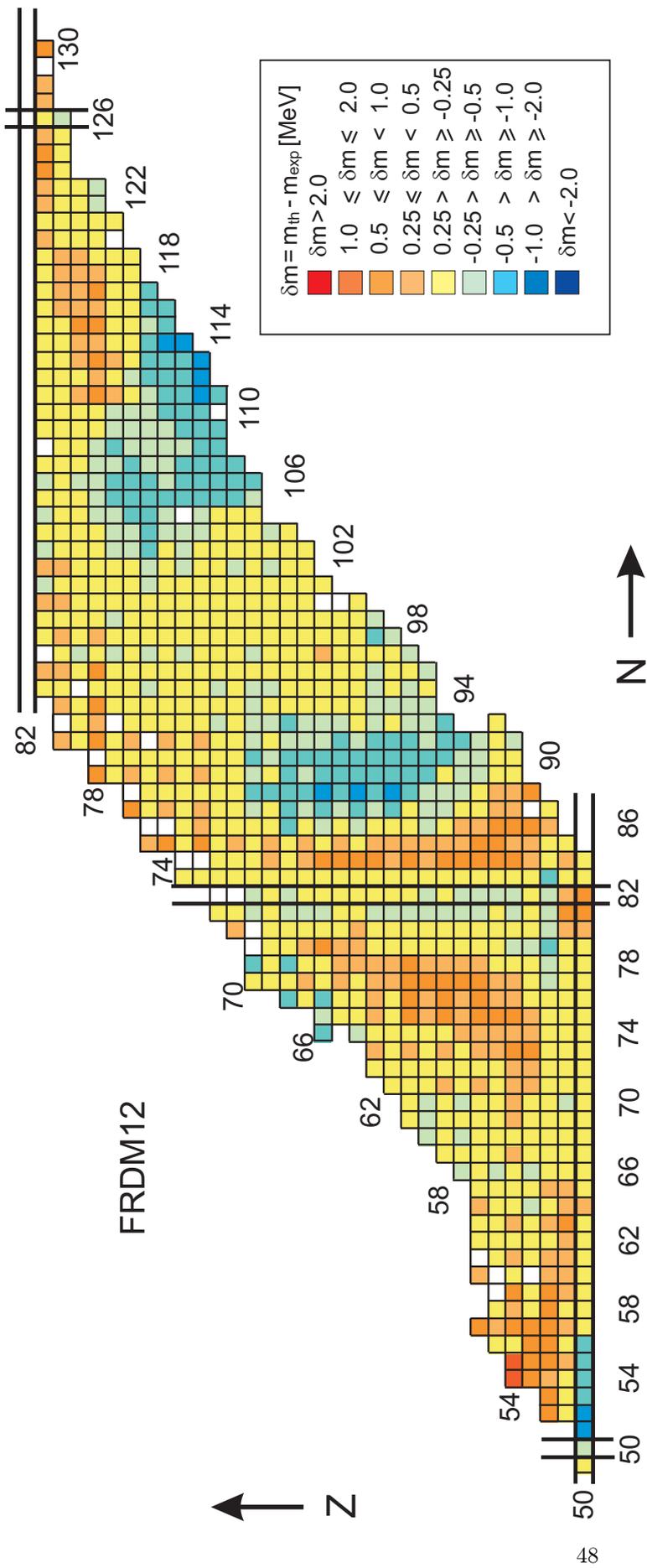

Fig. 45:

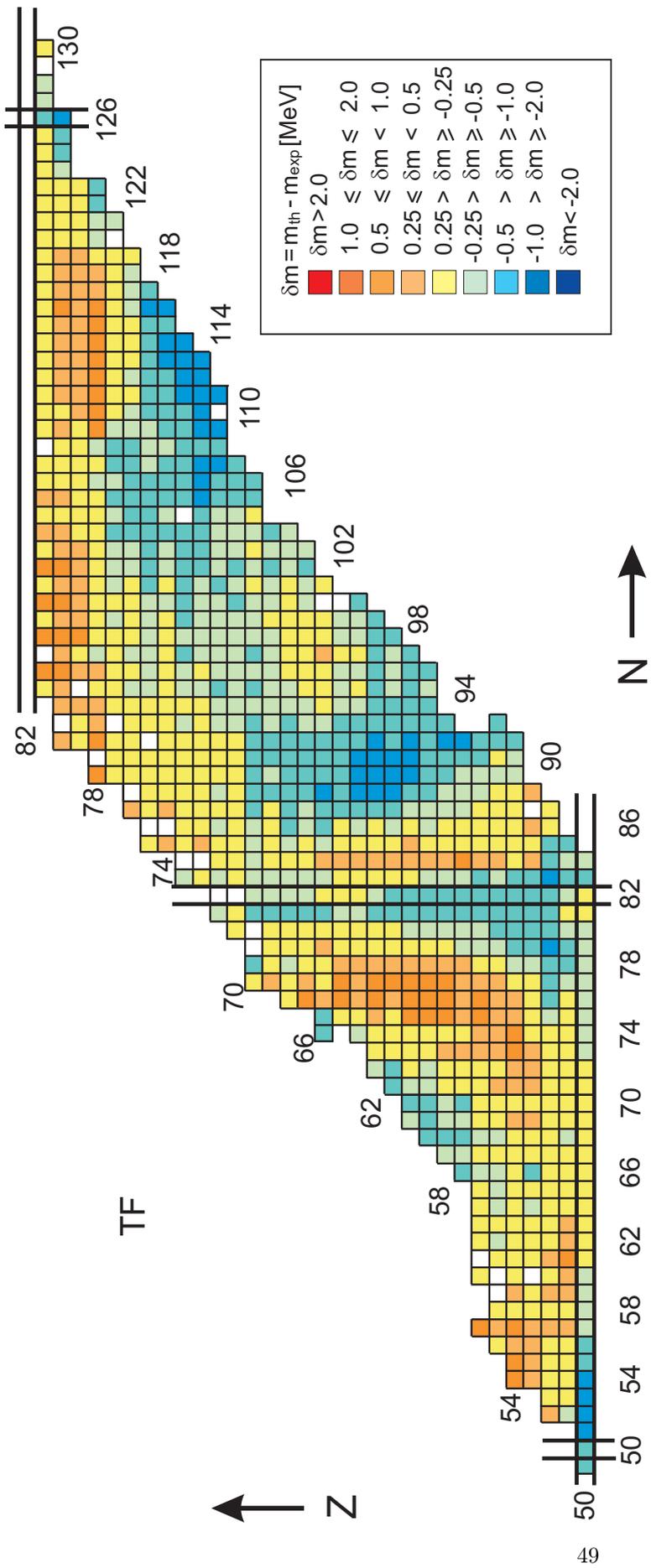

Fig. 46:

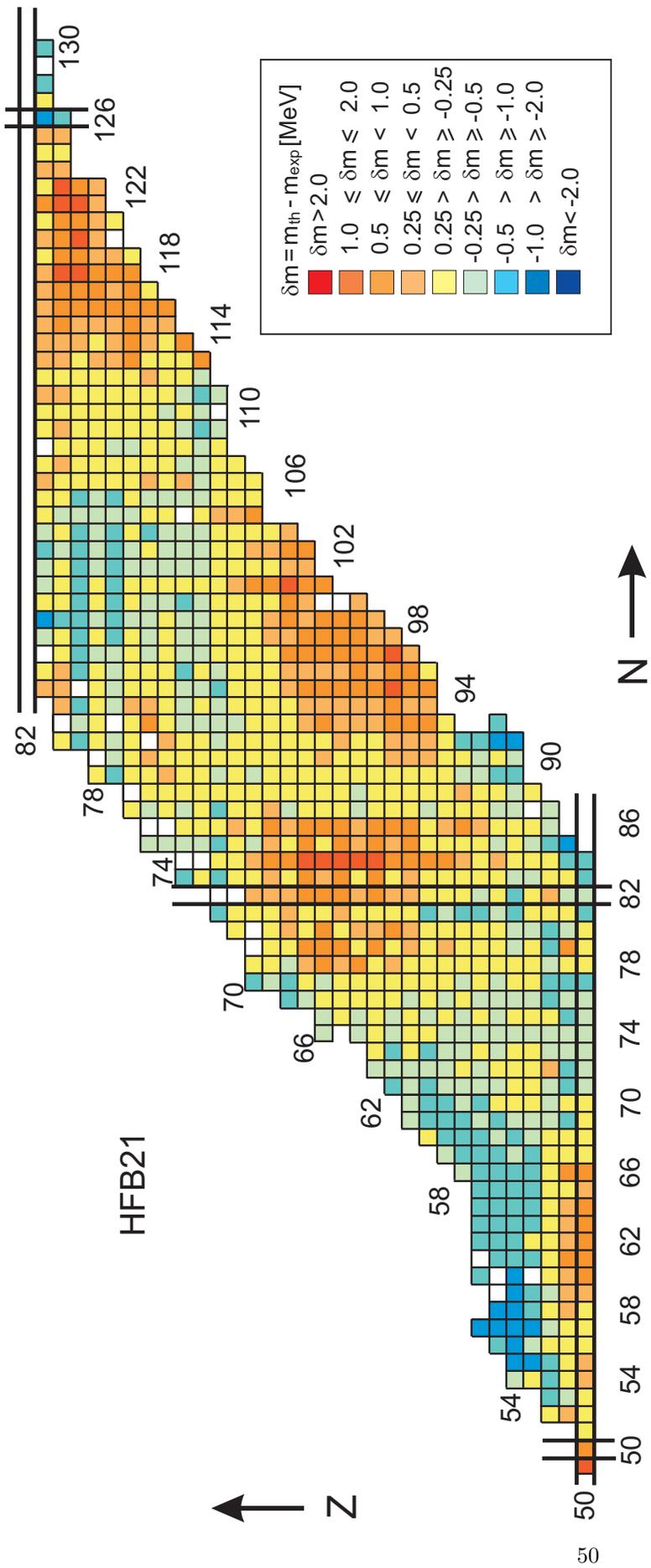

Fig. 47:

GHFB

Fig. 48:

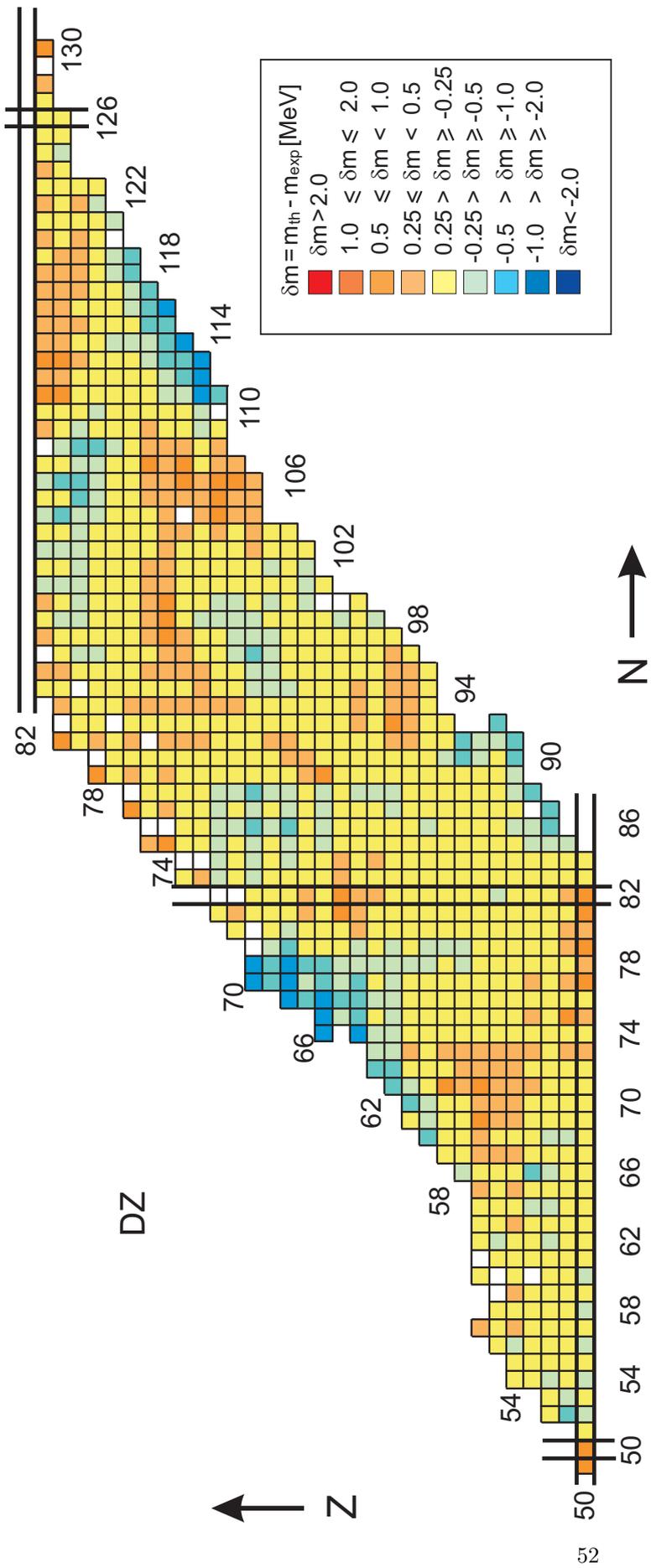

Fig. 49:

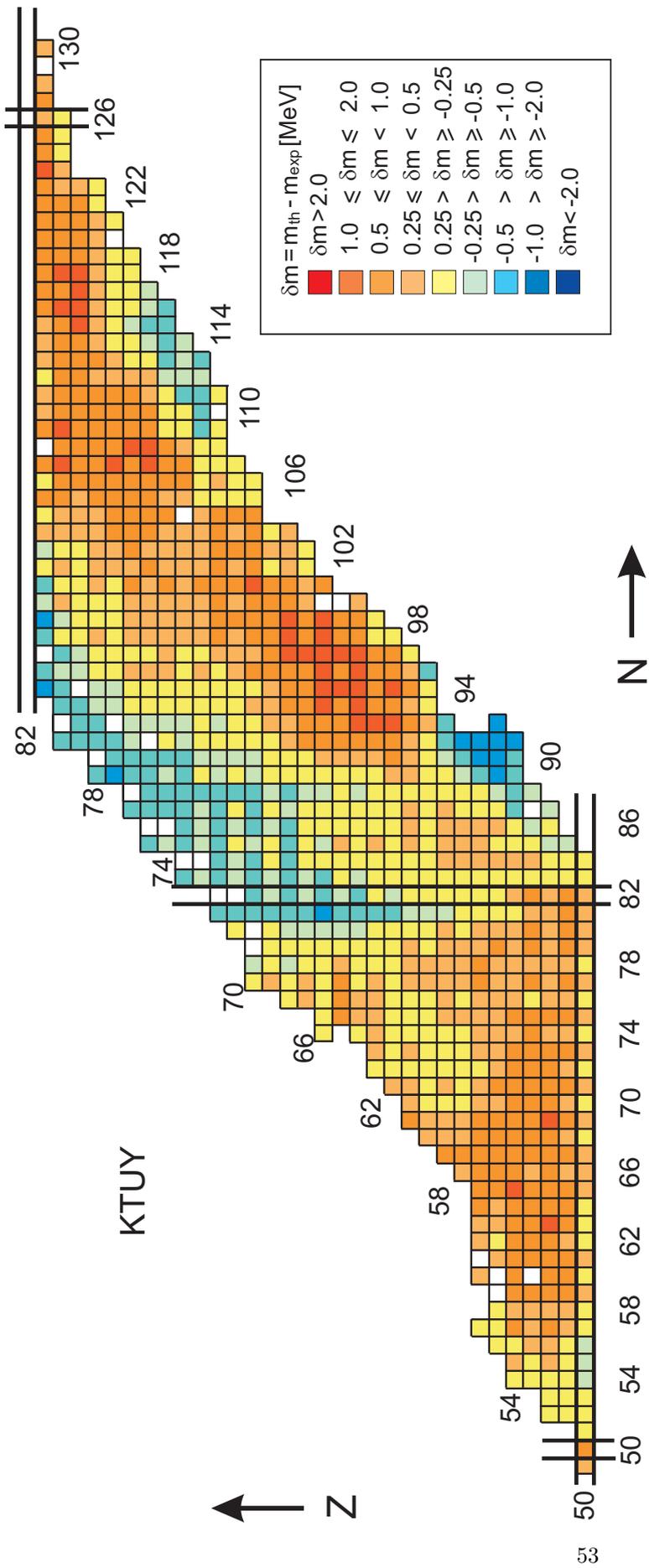

Fig. 50:

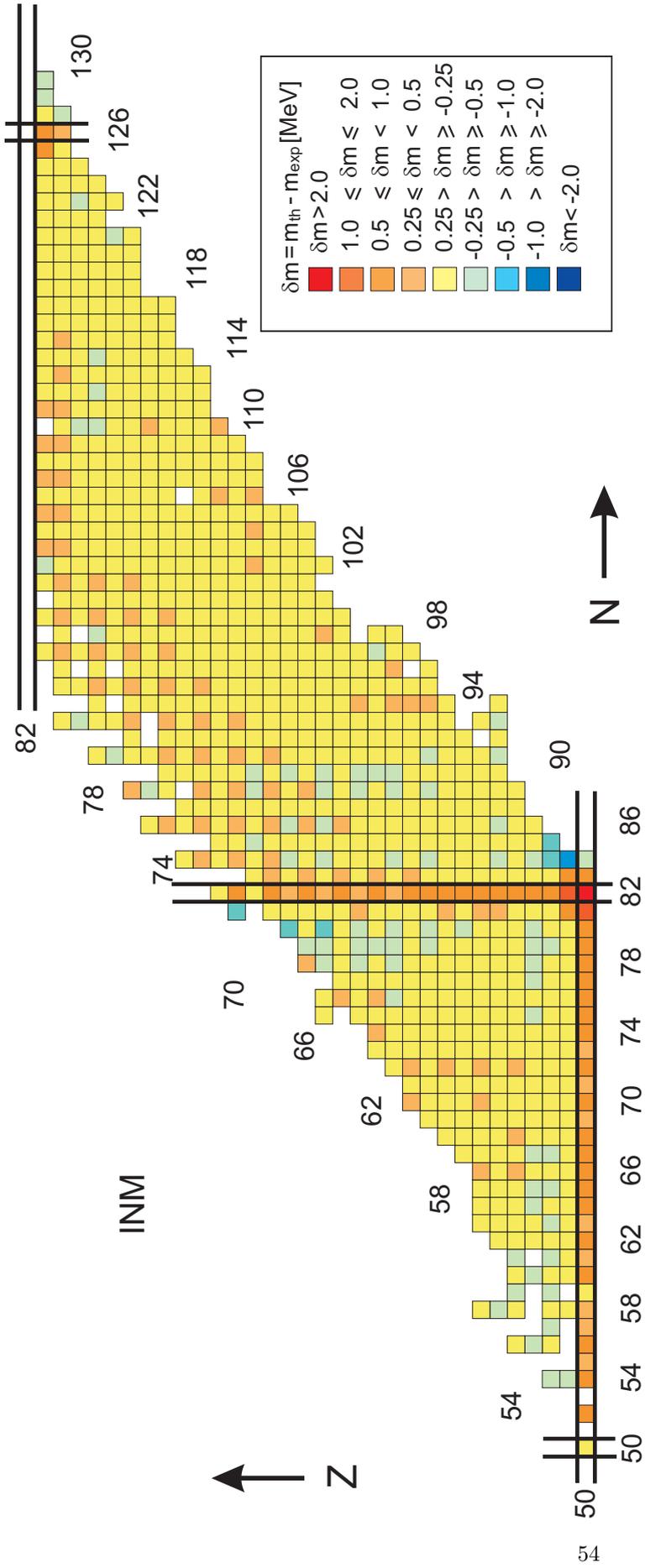

Fig. 51:

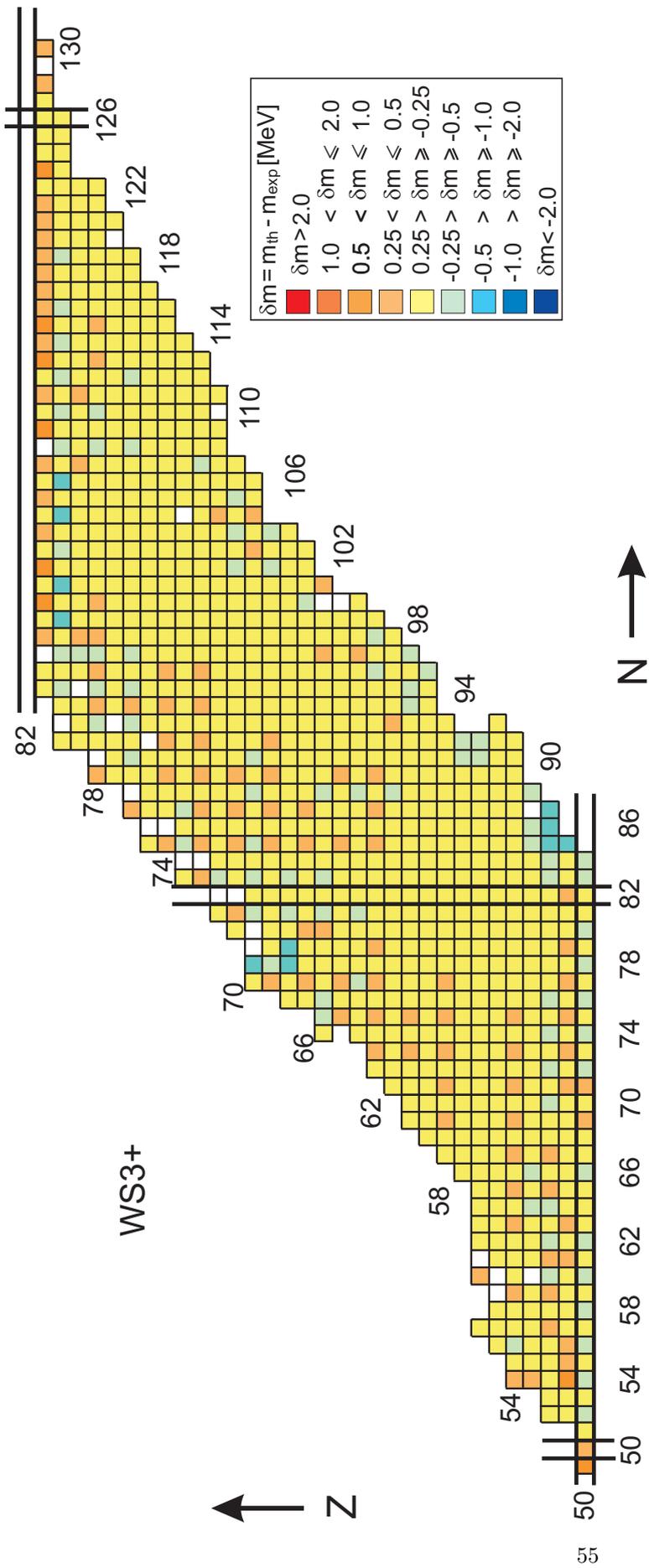

Fig. 52:

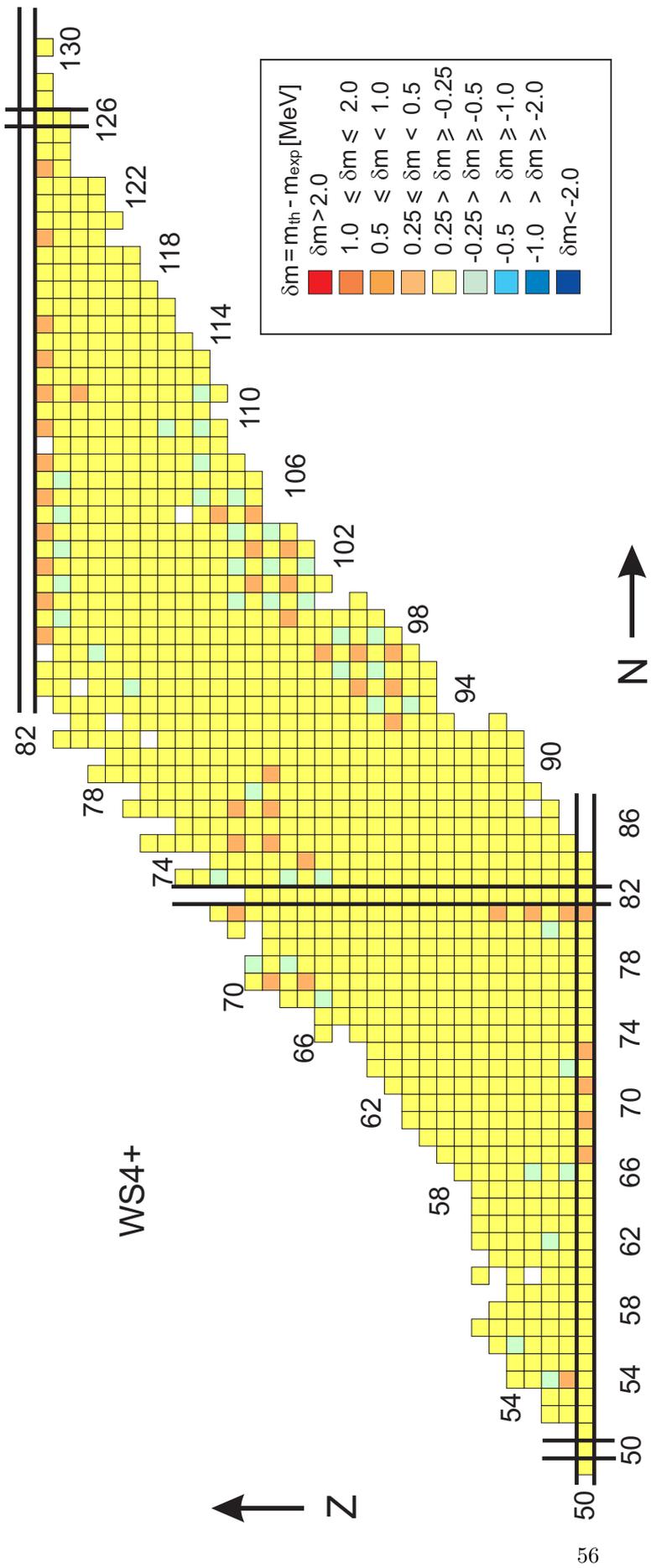

Fig. 53:

*5.4. Heavy nuclei*

The tendency of the increasing the accuracy of the description of nuclear masses with increasing mass of nuclei is further observed.

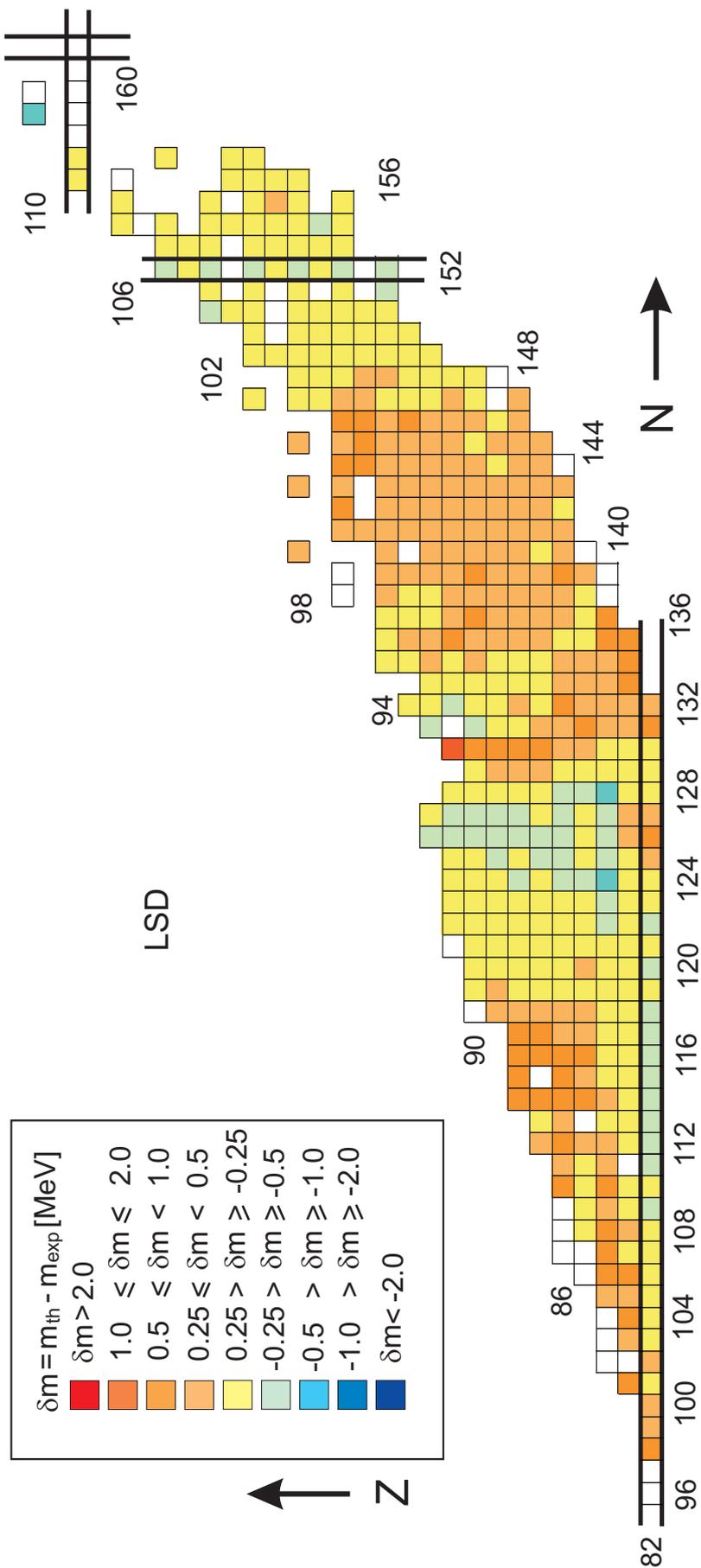

**Fig. 54:**

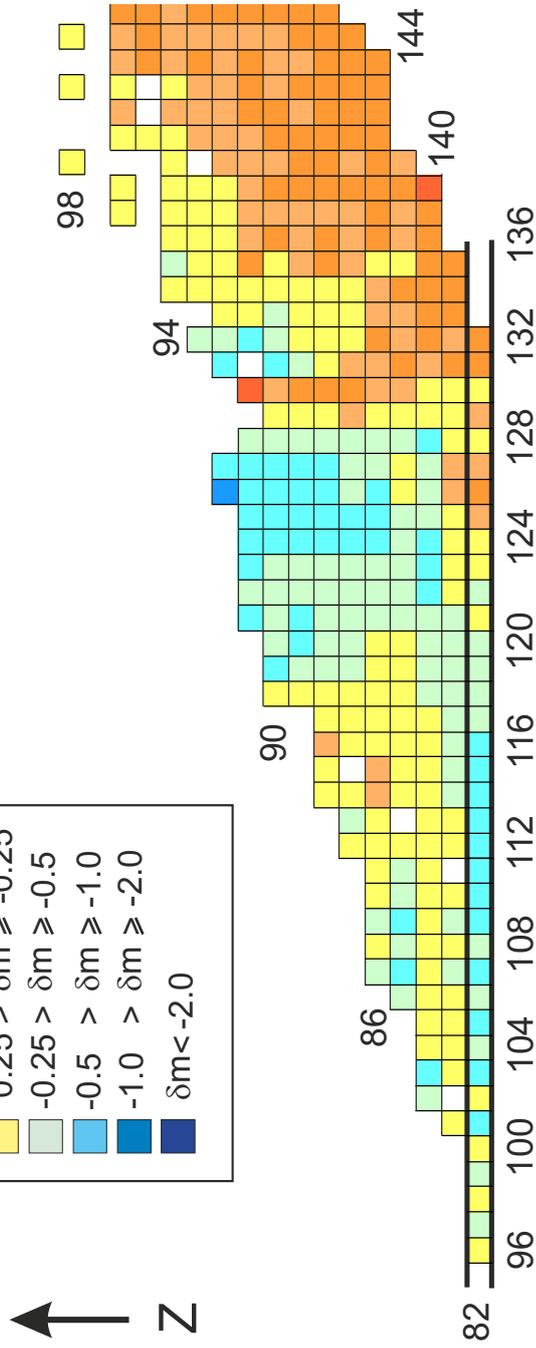

Fig. 55:

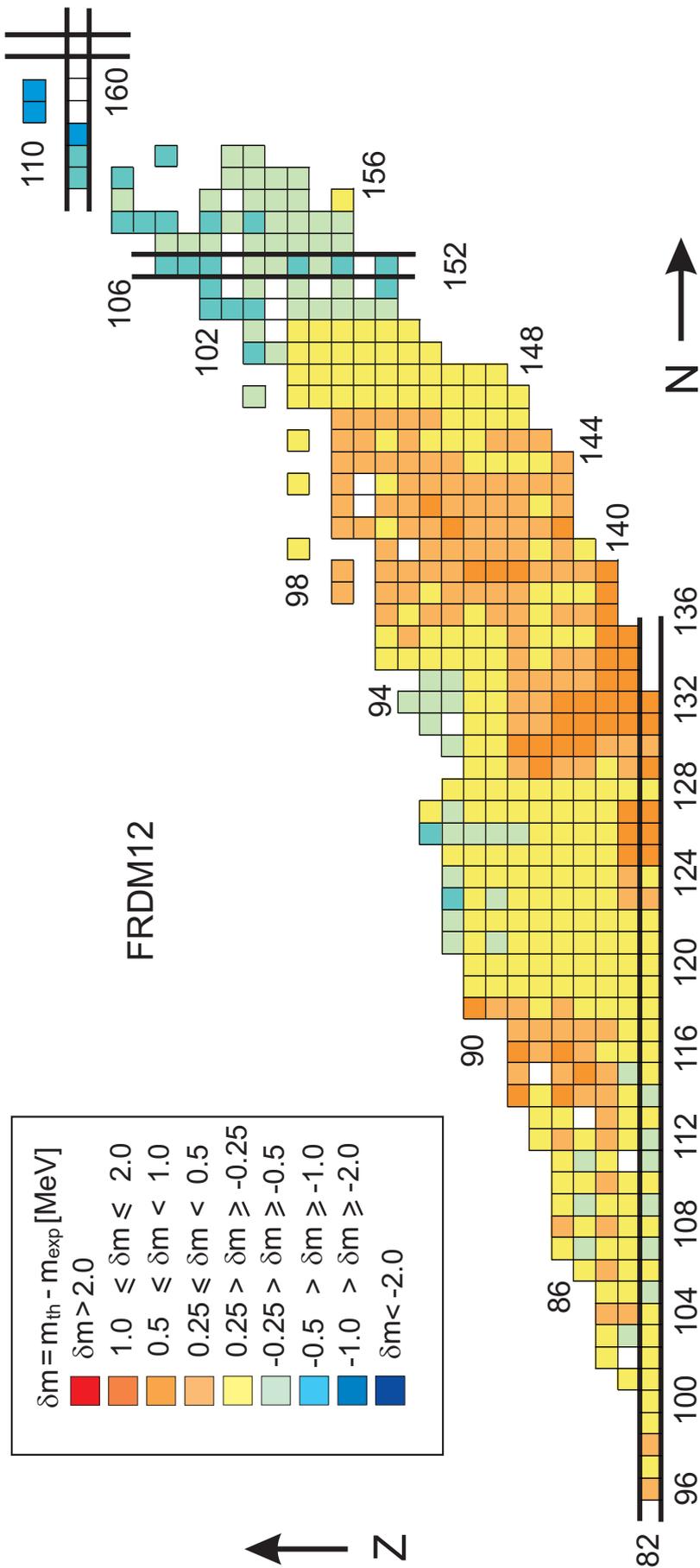

**Fig. 56:**

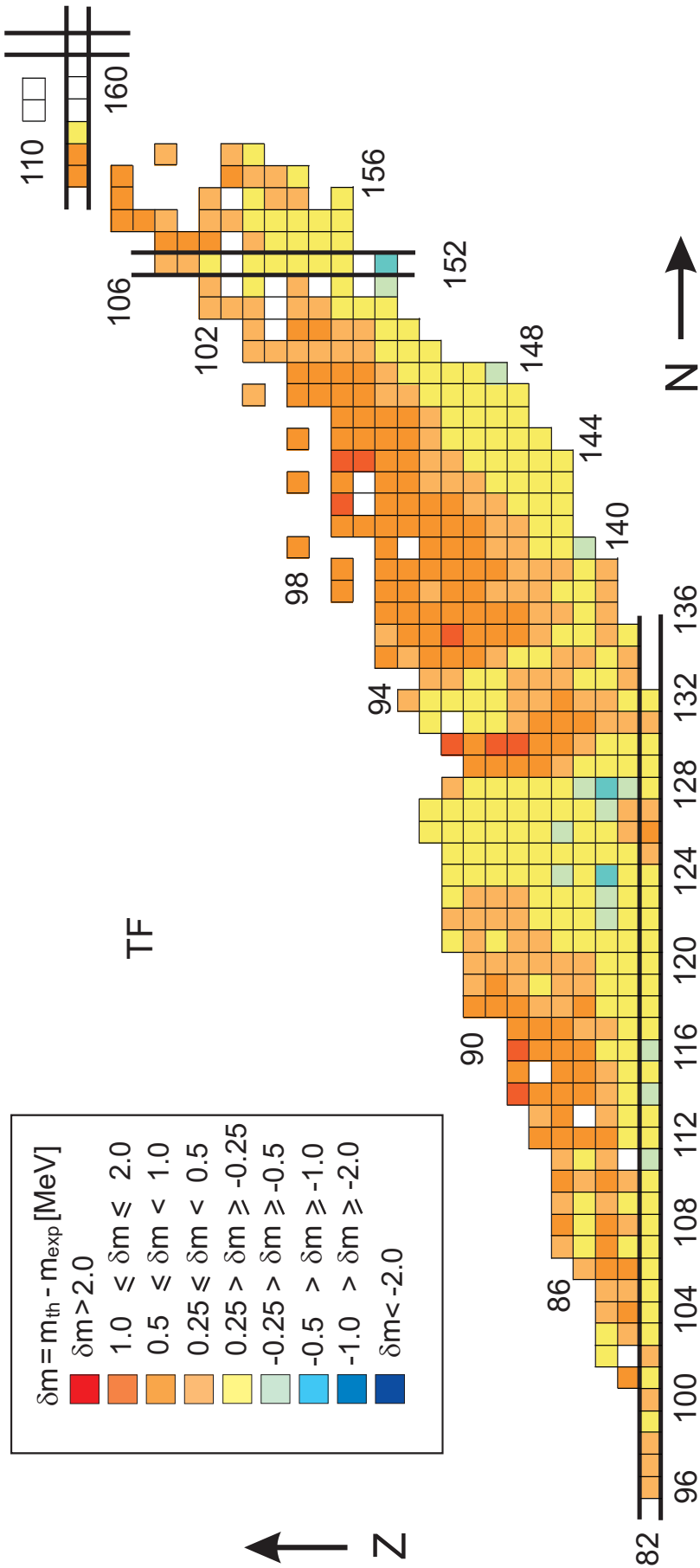

**Fig. 57:**

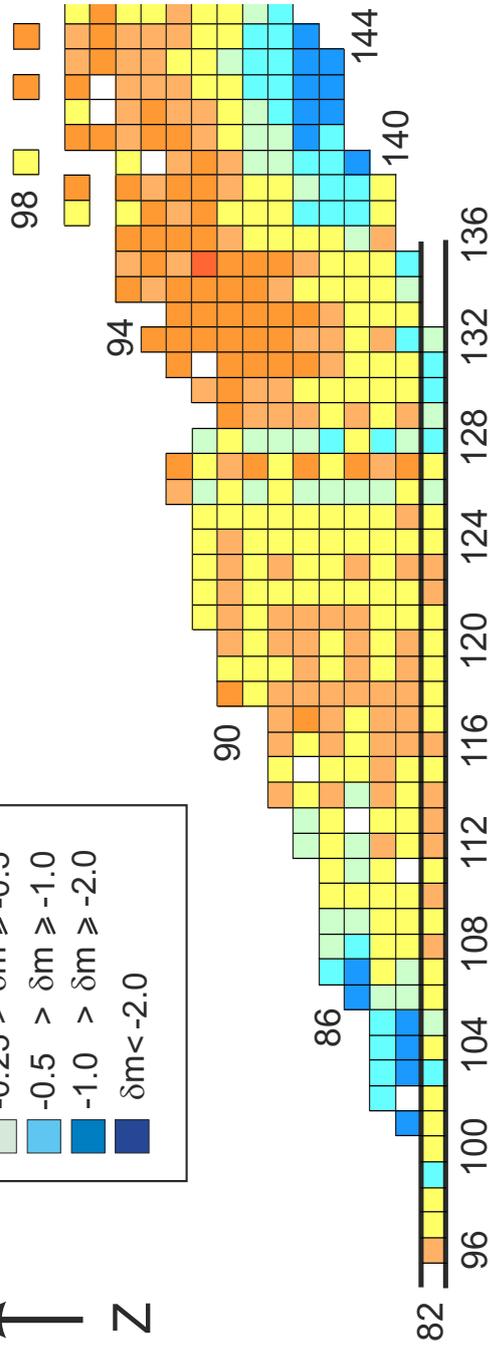
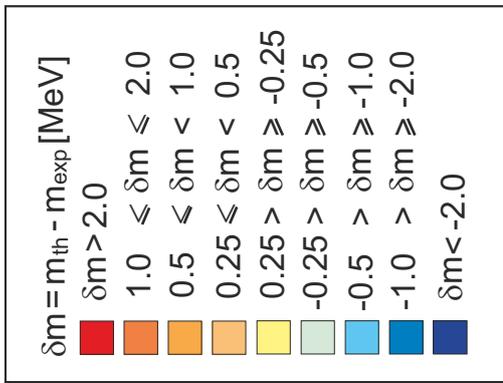

Fig. 58:

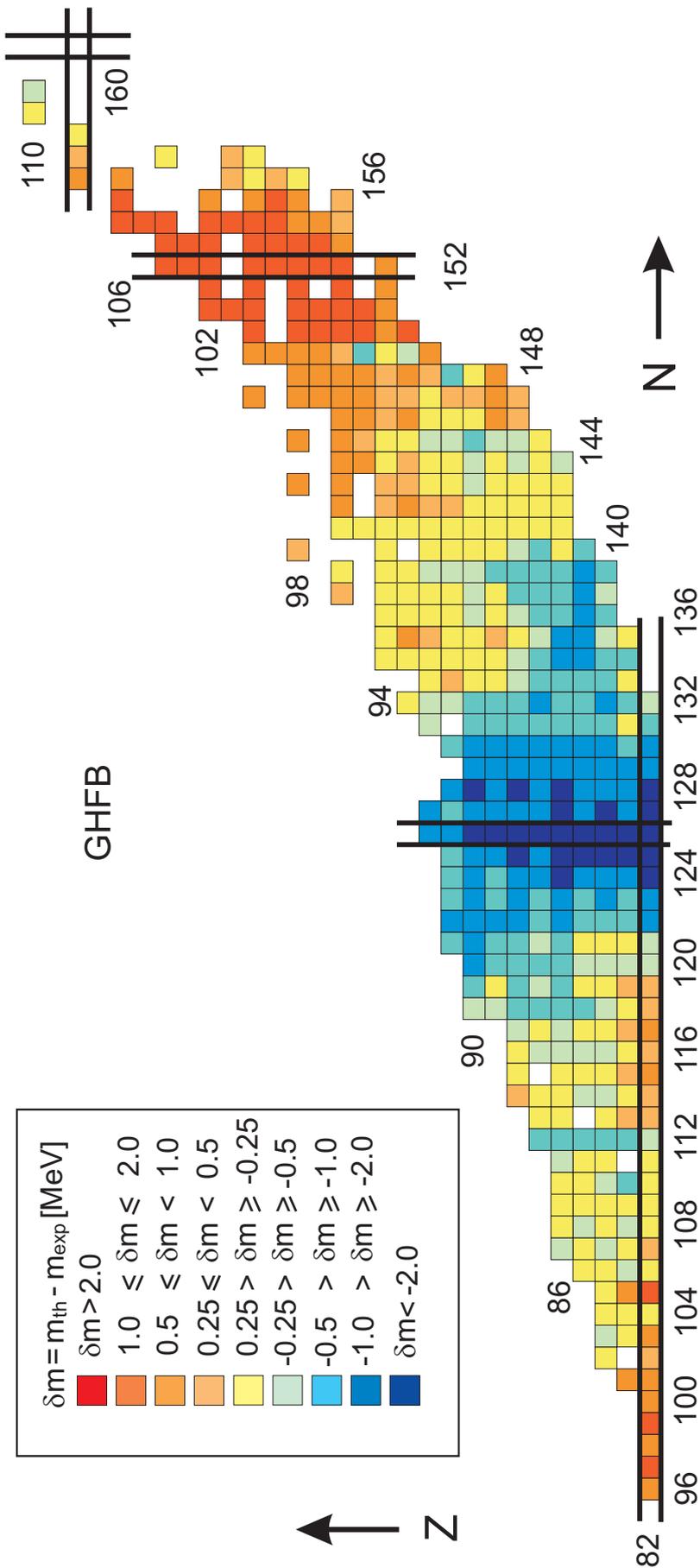

Fig. 59:

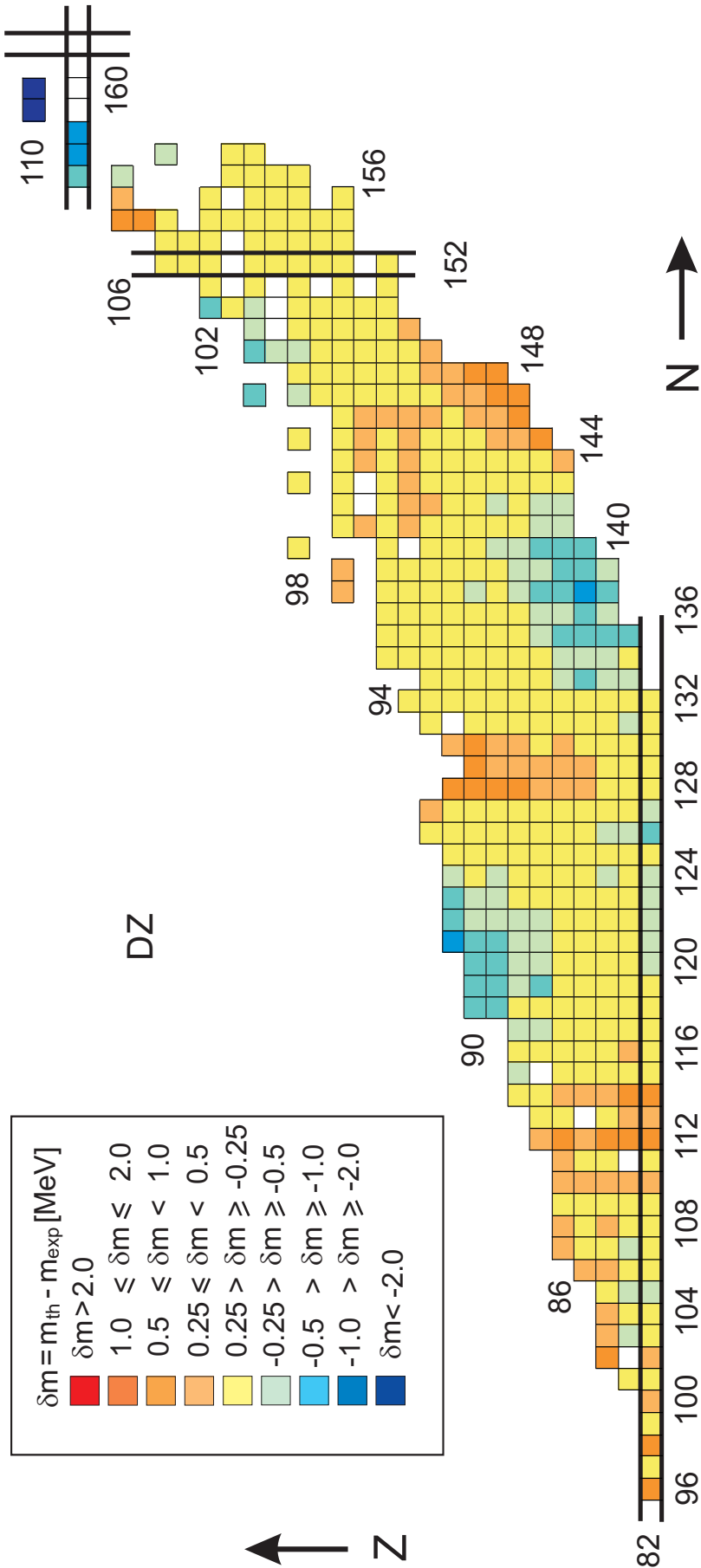

**Fig. 60:**

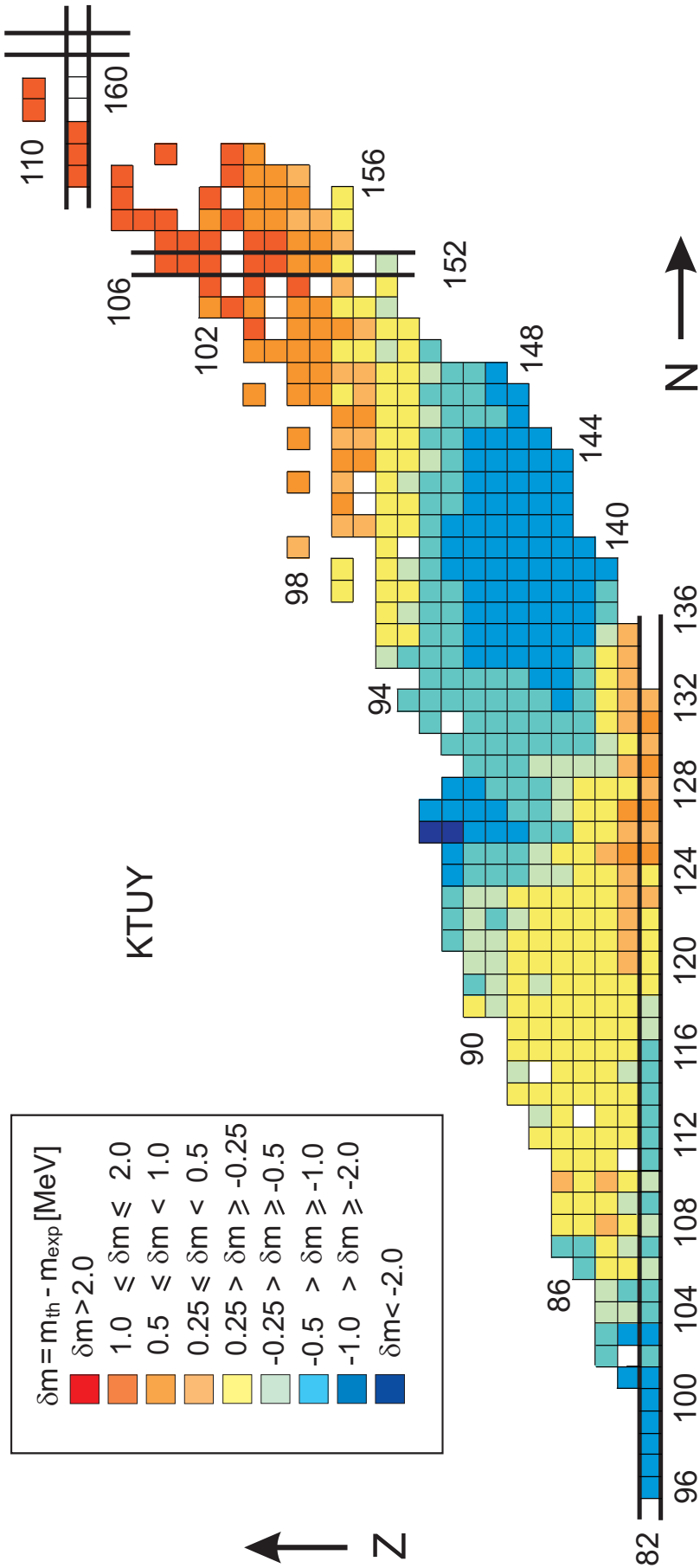

**Fig. 61:**

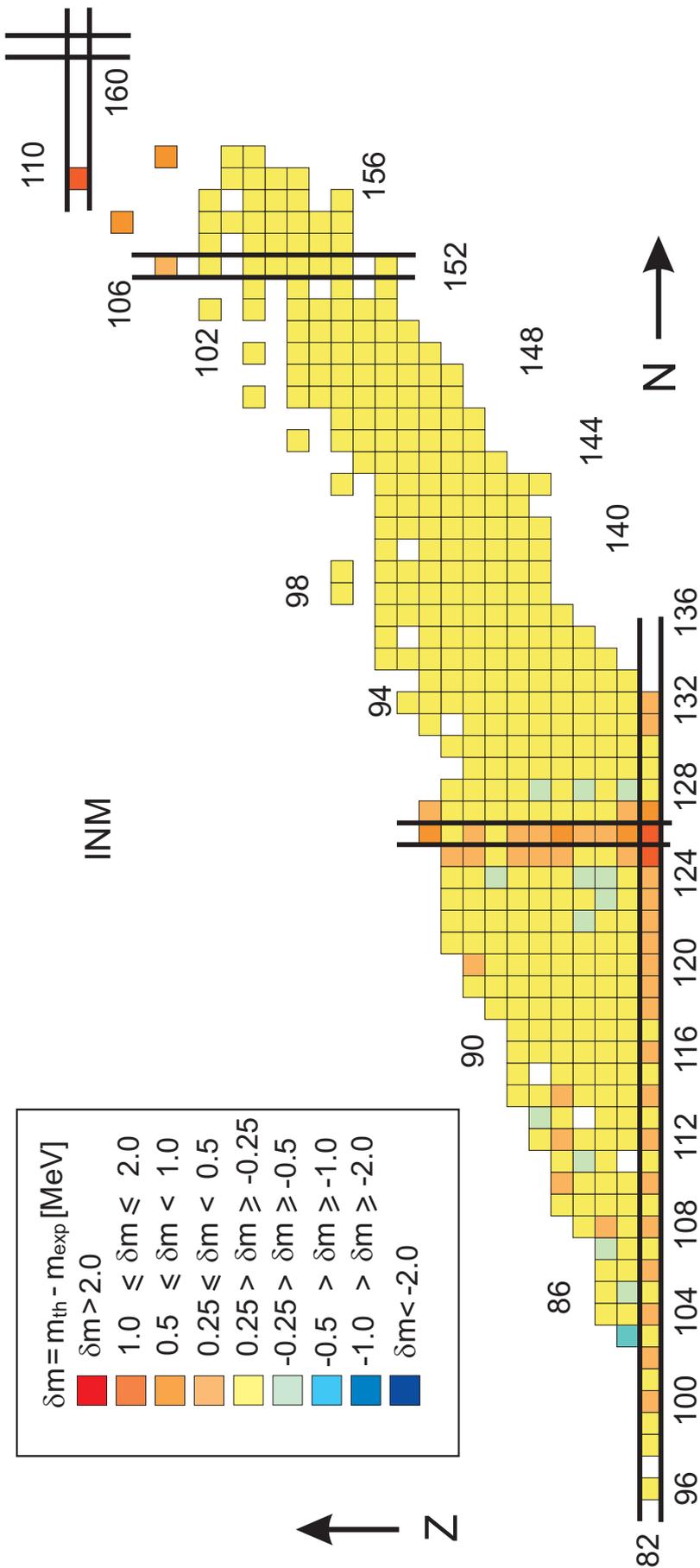

**Fig. 62:**

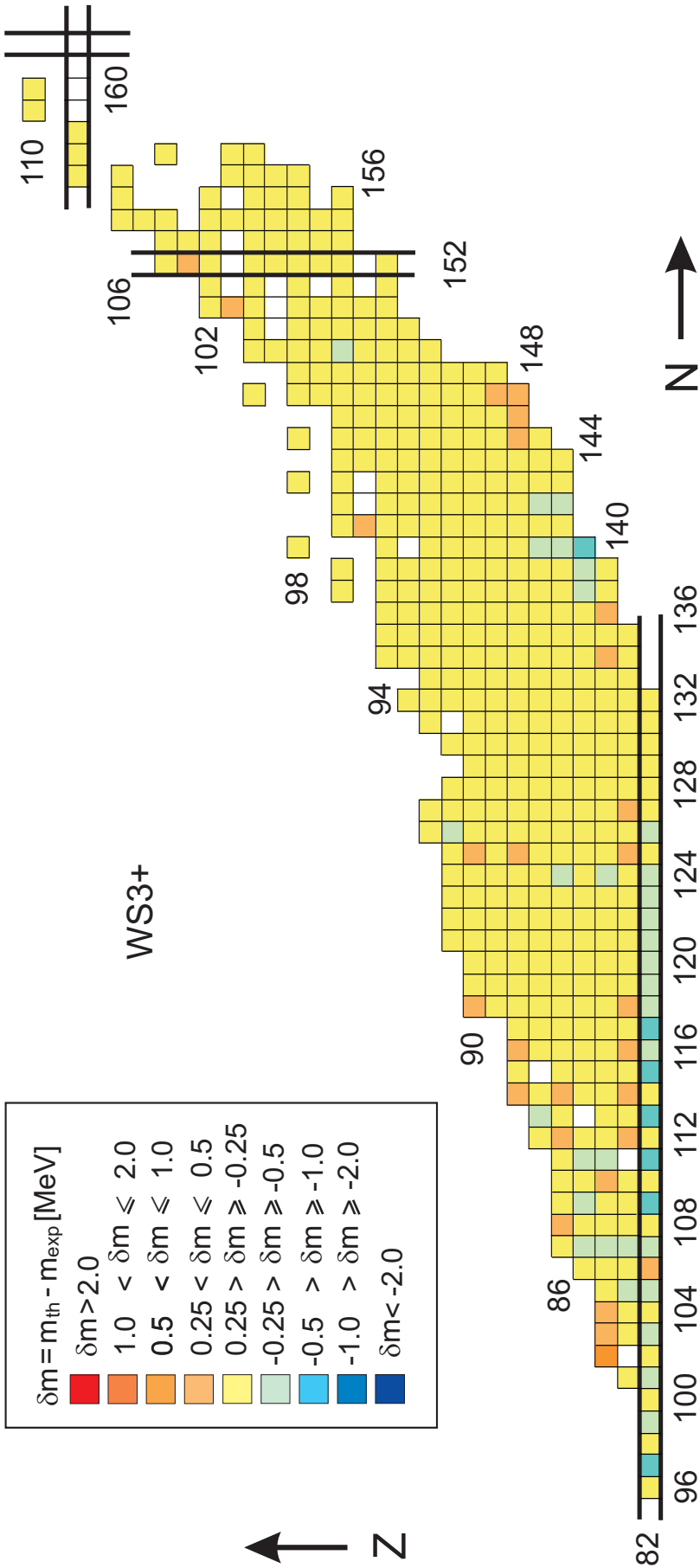

**Fig. 63:**

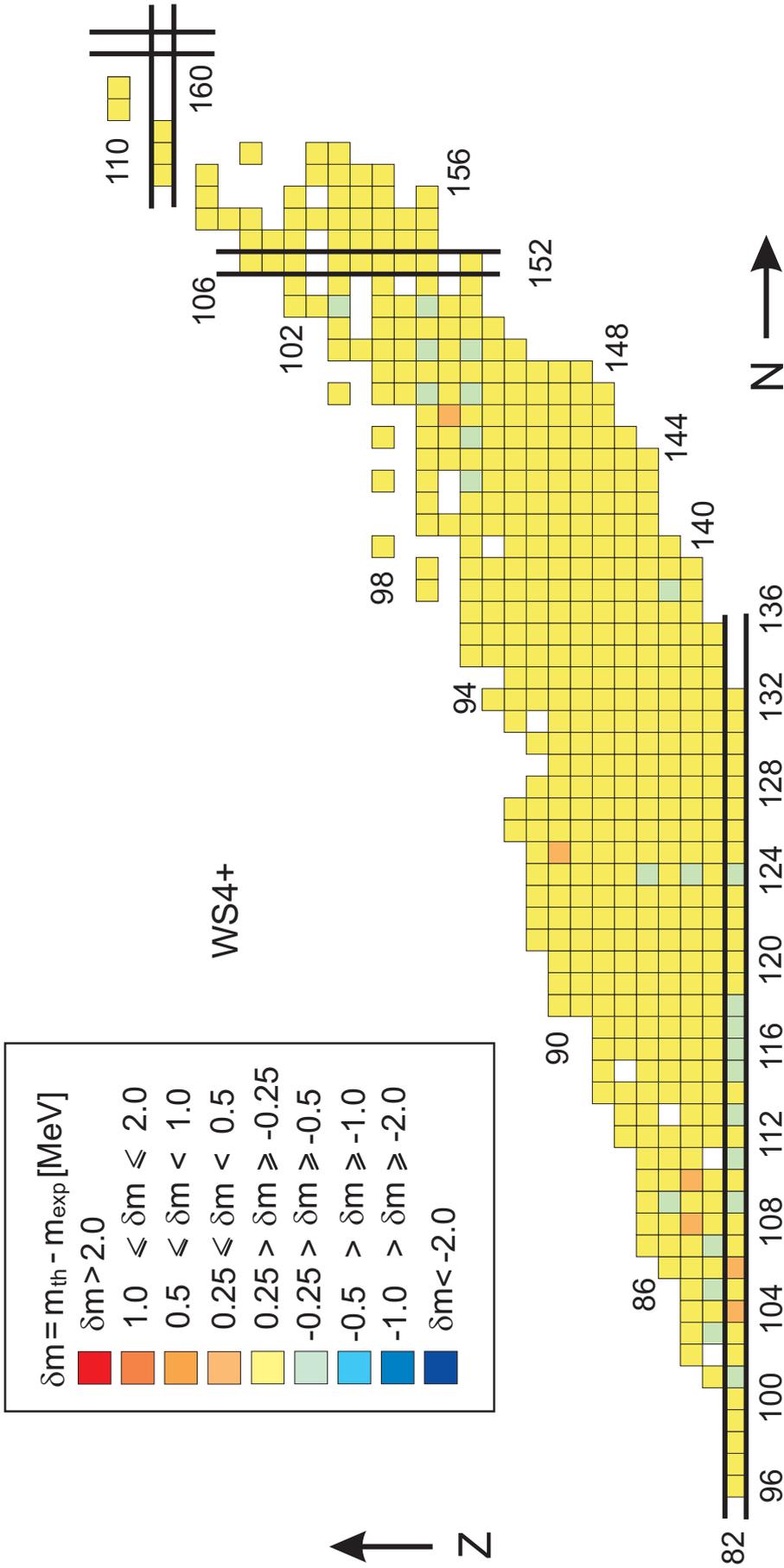

Fig. 64:

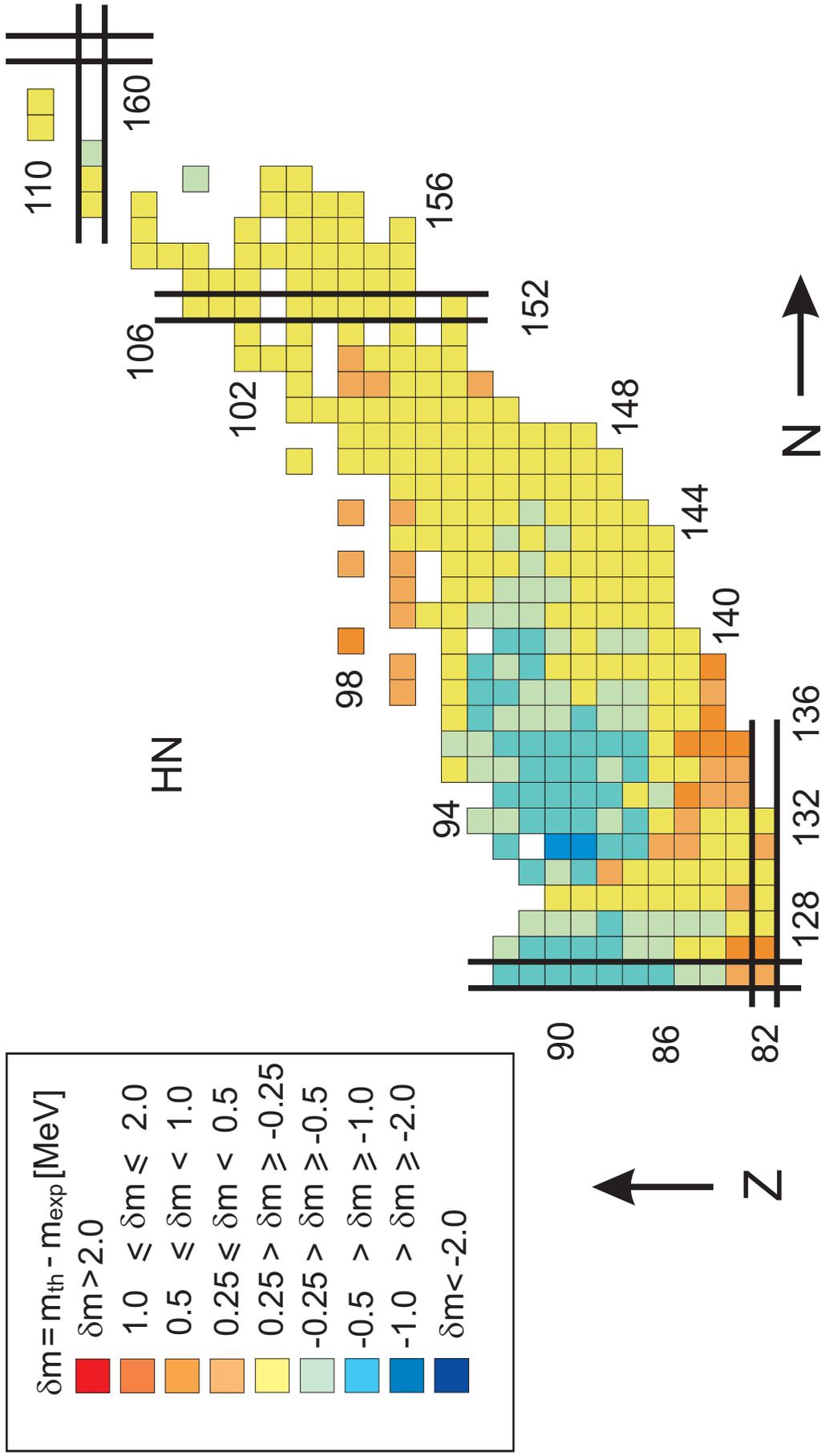

**Fig. 65:**

# 6. Summary and conclusions

The accuracy of description of measured nuclear masses by nuclear-mass models has been studied. Twelve models of various kinds have been considered, eleven of the global character and one local model specially adapted to description of heavy nuclei. To improve the illustration of the accuracy of the description, the global region of nuclei has been divided to four subregions. This allowed one to study and present the results of the accuracy for each nucleus separately, without any averaging. The results have been presented in the form of colored maps, large enough to be easily and accurately read.

Besides the accuracy of the studied models, also their predictive power has been considered.

The following conclusions may be drawn from the study:

(1) The accuracy of description of mass strongly depends on a nuclear-mass model and on the region of nuclei to which it is applied.

(2) For most of the models, the accuracy increases with the increase of mass of nuclei. This may be connected with the better mean field for heavier nuclei.

(3) A better accuracy of description of mass of deformed nuclei than of spherical ones (e.g. magic nuclei) is observed. This may be connected with more correlations (mixture of single-particle configurations) in deformed nuclei than in the spherical ones.

(4) The best accuracy of description of mass in all subregions of nuclei is obtained by the recent two Chinese models WS3+ and WS4+. Concerning, however, the WS4+ model, the high accuracy of it may be due to large extent to the adjustment of it to the recent evaluation of measured masses, while the other models have been fitted to previous evaluations. A relatively good accuracy is also obtained by the INM and DZ models, as can be seen in Table A and in the respective maps.

(5) Generally, no clear, strong correlation between the accuracy of description of already known masses by a given model and its predictive power for new masses is observed. Still, such correlation is obtained in more cases of models and nuclear regions for the macro-micro models than for the other approaches. This might be connected with the presence of the macroscopic part (usually liquid drop) in these approaches, which is a physically good model for a number of nuclear properties, among them mass, averaged over microscopic effects.

++++++++++++++++++ Highlights ++++++++++++++++++++++

- Accuracy of mass description by 12 different models
- Dependence of accuracy on an investigated model
- Dependence of accuracy on the region of nuclei
- Accuracy of a model for each of 2353 nuclei separately
- Predictive power of a model
- Correlation between accuracy and predictive power of a model